\documentclass[]{jfm}

\usepackage{graphicx,color,amsmath,amssymb}
\usepackage{natbib}
\usepackage{newtxtext}
\usepackage{newtxmath}
\usepackage{multirow}
\usepackage{hyperref}
\hypersetup{
	colorlinks = true,
	urlcolor   = blue,
	citecolor  = black,
}

\usepackage{comment}

\usepackage{cleveref}
\crefname{section}{section}{sections}
\crefname{appsec}{appendix}{appendices}
\crefname{subsection}{subsection}{subsections}
\crefname{figure}{figure}{figures}
\crefname{table}{table}{tables}
\crefname{equation}{}{}
\Crefname{section}{Section}{Sections}
\Crefname{appsec}{Appendix}{Appendices}
\Crefname{subsection}{Subsection}{Subsections}
\Crefname{figure}{Figure}{Figures}
\Crefname{table}{Table}{Tables}

\newcommand{\RomanNumeralCaps}[1]
\linenumbers

\newcommand\solidrule[1][15pt]{\rule[0.5ex]{#1}{1pt}}
\newcommand\dashedrule{\mbox{%
		\solidrule[3pt]\hspace{3pt}\solidrule[3pt]\hspace{3pt}\solidrule[3pt]}}

\newcommand{\beq}{\begin{equation}}
	\newcommand{\eeq}{\end{equation}}
\newcommand{\vu}{\mathbf{u}}
\newcommand{\vvv}{\mathbf{v}}
\newcommand{\vw}{\mathbf{w}}
\newcommand{\vx}{\mathbf{x}}

\newcommand{\vxhat}{\hat{\mathbf{x}}}
\newcommand{\vz}{\mathbf{z}}
\newcommand{\vzhat}{\hat{\mathbf{z}}}

\newcommand{\vF}{\mathbf{F}}
\newcommand{\curl}{\nabla\times}
\newcommand{\funcs}{\mathcal{S}}
\newcommand{\funcr}{\mathcal{R}}
\newcommand{\funcq}{\mathcal{Q}}

\newcommand{\funce}{\mathcal{E}}
\newcommand{\funcf}{\mathcal{F}}
\newcommand{\funch}{\mathcal{H}}
\newcommand{\ccurl}{\nabla\times\nabla\times}
\newcommand{\pprime}{\prime\prime}
\newcommand{\lap}{\nabla^2}
\newcommand{\dx}{{\rm d}\vx}
\newcommand{\dt}{{\rm d}t}
\newcommand{\ddt}{\tfrac{\rm d}{\dt}}
\newcommand{\Hdot}{\dot{\mathcal{H}}}
\newcommand{\wn}{\mathbf{k}}
\newcommand{\dotw}{\dot{\vw}}

\newcommand{\eps}{\varepsilon}

\title{Bounds on dissipation in three-dimensional planar shear flows: reduction to two-dimensional problems}

\author{Farid Rajkotia-Zaheer
\corresp{\email{faridrajkotiazaheer@uvic.ca, goluskin@uvic.ca}} \and David Goluskin$\dag$}

\affiliation{Department of Mathematics and Statistics, University of Victoria, Victoria, BC, V8P 5C2, Canada}

\begin{document}
\maketitle

\begin{abstract}
Bounds on turbulent averages in shear flows can be derived from the Navier--Stokes equations by a mathematical approach called the background method. Bounds that are optimal within this method can be computed at each Reynolds number $\Rey$ by numerically optimizing subject to a spectral constraint, which requires a quadratic integral to be nonnegative for all possible velocity fields. Past authors have eased computations by enforcing the spectral constraint only for streamwise-invariant (\mbox{2.5-D}) velocity fields, assuming this gives the same result as enforcing it for three-dimensional (\mbox{3-D}) fields. Here we compute optimal bounds over \mbox{2.5-D} fields and then verify, without doing computations over 3-D fields, that the bounds indeed apply to \mbox{3-D} flows. One way is to directly check that an optimizer computed using \mbox{2.5-D} fields satisfies the spectral constraint for all \mbox{3-D} fields. A second way uses a criterion we derive that is based on a theorem of \citet{busse_property_1972} for energy stability analysis of models with certain symmetry. The advantage of checking this criterion, as opposed to directly checking the \mbox{3-D} constraint, is lower computational cost and natural extrapolation of the criterion to large $\Rey$. We compute optimal upper bounds on friction coefficients for the wall-bounded Kolmogorov flow known as Waleffe flow, and for plane Couette flow. This requires lower bounds on dissipation in the first model and upper bounds in the second. For Waleffe flow, all bounds computed using \mbox{2.5-D} fields satisfy our criterion, so they hold for \mbox{3-D} flows. For Couette flow, where bounds have been previously computed using \mbox{2.5-D} fields by \citet{plasting_improved_2003}, our criterion holds only up to moderate $\Rey$, so at larger $\Rey$ we directly verify the \mbox{3-D} spectral constraint. Over the $\Rey$ range of our computations, this confirms the assumption by \citeauthor{plasting_improved_2003} that their bounds hold for \mbox{3-D} flows.
\end{abstract}


\section{Introduction}
\label{sec:intro}

Some of the most fundamental questions about turbulent fluid flows concern space- and time-averaged quantities, such as mean dissipation or transport, and how these quantities scale with control parameters. At parameter values that are accessible to laboratory experiments or direct numerical simulations, mean quantities can be estimated by averaging over a finite-time flow. A different and complementary approach is to mathematically derive upper or lower bounds on infinite-time averages directly from the governing equations. Most bounds of this type have been derived using the so-called background method, which was first applied to the Navier--Stokes equations by \citet{doering_energy_1992}. For an overview of the method see \citet{chernyshenko_relationship_2022} and \citet{fantuzzi_background_2021}. 

The background method lets the time-dependent governing equations be replaced by variational problems, in which integrals are maximized or minimized over time-independent velocity fields. For instance, upper bounds on infinite-time-averaged dissipation at fixed parameter values can be formulated, roughly speaking, as
\begin{equation}
\text{mean dissipation} \le \min_{\zeta}\max_{\vw}\funcq[\vw;\zeta],
\label{eq: intro}
\end{equation}
where $\funcq[\vw;\zeta]$ is a spatial integral whose integrand depends quadratically on an incompressible velocity field $\vw$ and linearly on a `background profile' $\zeta$. In the case of lower bounds, the inner problem is a minimization over $\vw$ while the outer one is a maximization over $\zeta$. A precise version of \cref{eq: intro} for planar shear flows is derived in \cref{sec: background method}.

The inner maximum in \cref{eq: intro} is an upper bound on dissipation for any admissible choice of~$\zeta$. For some $\zeta$ this bound is infinity, but for other $\zeta$, is finite. The outer minimization over bounds in \cref{eq: intro} gives the optimal bound within the background method framework. Optimal bounds generally cannot be found analytically, but they have been computed numerically for a few fluid systems \citep{plasting_improved_2003,fantuzzi_optimization_2017,fantuzzi_bounds_2018,fantuzzi_background_2021}. As with direct numerical simulation of fluids, computation of optimal bounds is possible when parameters are fixed to values that are not too extreme, so that the required spatial resolution is not too fine. Most applications of the background method have instead derived suboptimal bounds analytically, which can give bounds applying at all parameter values, including with explicit parameter dependence. Such analytical results are derived by choosing relatively simple $\zeta$ that are suboptimal, then upper-bounding the maximum over $\vw$ rather than computing it exactly. 

For bounds like \cref{eq: intro} to hold for three-dimensional (\mbox{3-D}) flows, the inner maximization generally must be over \mbox{3-D} incompressible velocity fields. Maximizing over a smaller class of $\vw$ can make the maximum smaller and thus is not guaranteed to give an upper bound for \mbox{3-D} flows. A crucial exception occurs when one can show mathematically that a maximum over \mbox{3-D} velocity fields coincides with a maximum over a class of lower-dimensional velocity fields, in which case, the maximum in \cref{eq: intro} need only be taken over the smaller class. This dimension reduction is significant for numerical computations of optimal bounds, which becomes much easier, and it may also improve analytical bounds. The present work concerns how to solve the min--max problem~\cref{eq: intro} over lower-dimensional velocity fields and then show \emph{a posteriori} that the inner maximum would be the same over \mbox{3-D} fields, thus avoiding \mbox{3-D} computations.

Here we consider planar shear flows that are bounded by two parallel walls and are periodic in the other two directions. Such flows may be sustained by boundary conditions, body forcing, or both, and we assume that the governing model admits a laminar flow in a single direction. The laminar flow's direction is called the streamwise direction, the other periodic direction is called spanwise, and the bounded direction is called wall-normal. Our particular focus is on models whose governing equations are symmetric under 180 degree rotation around a spanwise axis. The most prominent models in this family are plane Couette flow and any wall-bounded Kolmogorov flows with forcing profiles that are odd about the mid-plane, including the half-period sinusoidal forcing sometimes called Waleffe flow \citep{waleffe_self-sustaining_1997}. Our main theoretical result is a criterion that applies only to shear flow models with such symmetry. In addition to fully \mbox{3-D} velocity fields, we will consider fields in only the wall-normal and streamwise directions, meaning there is neither flow nor variation
in the spanwise direction; these will be called two-dimensional (\mbox{2-D}). Fields that may be nonzero in all three components but do not vary in the streamwise direction will be called streamwise-invariant (\mbox{2.5-D}). Several past authors have assumed that a maximum over \mbox{2.5-D} fields in~\cref{eq: intro} coincides with a maximum over \mbox{3-D} fields. Our aim is to confirm such statements in specific cases without computing any extrema over \mbox{3-D} fields.

Plane Couette flow, which is driven by parallel relative motion of the walls, has been the most-studied application of the background method to the Navier--Stokes equations. This was the first model considered by \citet{doering_energy_1992,doering_variational_1994}, who derived an upper bound on dissipation. Normalizing the dissipation by its laminar value and by $\Rey$ gives a friction factor $\varepsilon$, for which the upper bounds become $\Rey$-independent at large $\Rey$. The bound $\varepsilon\le 2^{-7/2}\approx 0.0884$ of \citet{doering_energy_1992} is derived by choosing a simple suboptimal background profile $\zeta$ that depends on $\Rey$, then using functional inequalities to upper-bound the maximum in \cref{eq: intro} over \mbox{3-D} $\vw$ fields. Slightly improved analytical bounds were then derived by constructing closer-to-optimal $\zeta$ and upper-bounding the maximum over $\vw$ as sharply as possible \citep{gebhardt_rigorous_1995}. Still smaller bounds were found at various fixed $\Rey$ by numerically computing the inner maxima in \cref{eq: intro} and implementing the outer minimization over a restricted class of $\zeta$ \citep{nicodemus_variational_1997,nicodemus_background_1998,nicodemus_background_1998-1}. Finally, \citet{plasting_improved_2003} numerically carried out both the inner maximization and the outer minimization over the full class of $\zeta$ needed, constituting the first optimal bounds of the background method for any fluid flow. Their computed bounds on $\varepsilon$ approach a constant near 0.008553 as $\Rey\to\infty$, which remains the best known bound on dissipation for plane Couette flow. However, the analyses of \citet{plasting_improved_2003} and \citet{nicodemus_background_1998,nicodemus_background_1998-1} are not quite complete. In their computations, inner maximization in \cref{eq: intro} was generally carried out only over \mbox{2.5-D} $\vw$ fields. If this maximum is smaller than the maximum over \mbox{3-D} $\vw$, then it need not be a bound for \mbox{3-D} flows. (A bound for \mbox{2.5-D} flows alone is not needed since all \mbox{2.5-D} flows decay to the laminar state, as follows from energy stability analysis using only streamwise and wall-normal velocity components.) These authors maximized only over \mbox{2.5-D} $\vw$ because they conjectured that the maximum over fully \mbox{3-D} $\vw$ would give the same value. \Citeauthor{plasting_improved_2003} (personal communication) and \citet{nicodemus_variational_1997} confirmed this conjecture at a few modest parameter values by carrying out \mbox{3-D} computations, but they did not show it to be true in general.

We derive a criterion for shear flow models that, when it holds, implies coincidence of maxima in~\cref{eq: intro} over \mbox{2.5-D} fields and over \mbox{3-D} fields, and likewise for minima in the case of lower bounds. This criterion applies if the governing model has 180 degree rotational symmetry, but the flow need not be symmetric. We rely on a theorem of \citet{busse_property_1972} for energy stability of models with the rotational symmetry. Checking our criterion does not require extremizing over \mbox{3-D} fields. Instead, at fixed parameters, one must find the extremum of $\funcq$ and a related functional over \mbox{2.5-D} fields, extremize another related functional over \mbox{2-D} fields, and then check whether a ratio involving these three extrema is less than unity. In computational examples, the ratio asymptotes to a constant as $\Rey\to\infty$. If the asymptote is less than unity, this gives strong evidence that the coincidence of \mbox{2.5-D} and \mbox{3-D} extrema can be `extrapolated' to all~$\Rey$.

When our criterion is not useful because it is either false or inapplicable, one can instead check directly that maximizers in~\cref{eq: intro} are \mbox{2.5-D}. As explained later, the min--max problem~\cref{eq: intro} can be rewritten as a minimization subject to a so-called spectral constraint, which requires all eigenvalues of a certain linear eigenproblem to be nonnegative. This eigenproblem depends on the background profile $\zeta$ and can be solved independently for each pair of the streamwise and spanwise wavenumbers. After solving~\cref{eq: intro} over \mbox{2.5-D} $\vw$ and finding the optimal $\zeta$, one can check \emph{a posteriori} that no \mbox{3-D} fields violate the spectral constraint---that is, one can check the eigenvalues for nonzero streamwise wavenumbers. 

Here, we compute optimal background method bounds on dissipation for both Waleffe flow and Couette flow. In Waleffe flow, dissipation is maximized by the laminar state, so only lower bounds must be computed. In Couette flow the situation is reversed, and only upper bounds must be computed. For both flows we fix streamwise and spanwise periods of the domain and solve~\eqref{eq: intro} computationally up to moderate $\Rey$ values over \mbox{2.5-D} fields, then we verify that the maxima over \mbox{3-D} fields would coincide. In the case of Waleffe flow, the ratio used in our criterion asymptotes to a value well below unity, suggesting that~\eqref{eq: intro} will coincide for \mbox{2.5-D} and \mbox{3-D} fields at all $\Rey$ values. In the case of Couette flow, our computations roughly reproduce those of \citet{plasting_improved_2003}, but up to smaller $\Rey$ and with the spatial periods fixed. The ratio used in our criterion exceeds unity if $\Rey\gtrsim254$, so it cannot validate the assumption of \citeauthor{plasting_improved_2003} that maximizers are \mbox{2.5-D} at all $\Rey$. Instead, we directly check that the spectral constraint holds for \mbox{3-D} fields over our modest $\Rey$ range, which it does. To the extent that this spectrum can be extrapolated, it is consistent with the assumption of \citeauthor{plasting_improved_2003}. Additionally, we repeat the bounding computations for Couette flow with further constraints on $\zeta$ that guarantee our criterion will be satisfied, giving bounds for \mbox{3-D} flows that extrapolate to large $\Rey$ with a prefactor slightly worse than that of \citeauthor{plasting_improved_2003}.

This article is organized as follows. \Cref{sec: problem setup} formulates the background method for planar parallel shear flows, including three equivalent reformulations with the spectral constraint. \Cref{sec: main argument} derives our criterion for extrema over \mbox{2.5-D} and \mbox{3-D} fields to coincide in models with 180 degree rotational symmetry. \Cref{sec: computational} presents our computational applications to Waleffe flow and Couette flow, followed by conclusions in \cref{sec:conclusion}. The appendices provide details of certain arguments and computations, as well as an exposition of the proof of \citet{busse_property_1972} that underlies our own criterion.

\section{Four formulations of the background method for planar shear flows}
\label{sec: problem setup}

We consider an incompressible fluid flow bounded by two planar walls located at dimensionless coordinates $z=-1/2$ and $z=1/2$, where lengths have been scaled by the distance $d$ between the walls. Body forcing of the fluid and/or relative motion of the boundaries is assumed to point in only the $x$ direction, so that there exists a laminar flow in that direction. In the nomenclature of shear flows, the $x$ direction is streamwise, $y$ is spanwise, and $z$ is wall-normal. We assume the flow is periodic in the streamwise and spanwise directions with dimensionless periods of $\Gamma_x$ and $\Gamma_y$, respectively, so we let $-\Gamma_x/2\le x\le \Gamma_x/2$ and $-\Gamma_y/2\le y\le\Gamma_y/2$. The Navier--Stokes equations governing the dimensionless velocity vector $\vu(\vx,t)$ and pressure $p(\vx,t)$ are
\begin{equation}
\partial_t\vu+\vu\cdot\nabla\vu=-\nabla p+\tfrac{1}{\Rey}\lap\vu + f\vxhat,
\quad \nabla\cdot\vu=0,
\label{eq: NSE}
\end{equation}
where $\Rey=d\mathsf{U}/\nu$ is the Reynolds number, $\nu$ is the kinematic viscosity, $\mathsf{U}$ is a dimensional velocity defined either using the boundary conditions or body forcing, and time has been scaled by $d^2/\nu$. If there is body forcing it is in the streamwise direction $\vxhat$ and varies only in the wall-normal direction, so it takes the form $f(z)\vxhat$. The walls are impenetrable, meaning the wall-normal velocity $\vu\cdot\vzhat$ vanishes, and remaining boundary conditions fix the tangential components of either the velocity vector $\vu$ or the stresses $\partial_z\vu$. In the fixed-velocity case the boundary conditions are
\begin{equation}
u_1=C,~~u_2=0,~~u_3=0\quad \text{at}\quad z=\pm\tfrac12, 
\label{eq: no slip u}
\eeq
and in the fixed-stress case they are
\beq
\partial_z u_1=C,~~\partial_z u_2=0,~~u_3 = 0 \quad \text{at}\quad z=\pm\tfrac12,
\label{eq: stress free u}
\end{equation}
where $C$ can be any constant (including zero) and can be different at each boundary.

The configurations described previously admit a laminar solution $U(z)\vxhat$ to the governing equations~\eqref{eq: NSE}. Most derivations here are done in terms of the deviation $\vvv$ from the laminar state, which is defined by $\vu(\vx,t)=U(z)\mathbf{\hat{x}}+\vvv(\vx,t)$. Denoting the components of the deviation by $\vvv=(v_1,v_2,v_3)$, the evolution of $\vvv$ implied by \cref{eq: NSE} is
\begin{equation}\partial_t\vvv + \vvv\cdot\nabla\vvv + U\partial_x\vvv + U'v_3\hat{\vx} =- \nabla p + \tfrac{1}{\Rey}\lap\vvv,\quad \nabla\cdot\vvv=0,
	\label{eq: NSE v}
\end{equation}
where primes denote ordinary derivatives in $z$. Boundary conditions on the deviation $\vvv$ are homogeneous. At a boundary where $\vu$ satisfies the fixed-velocity condition~\cref{eq: no slip u}, the $\vvv$ condition is no-slip,
\begin{equation}
\vvv=\mathbf{0}\quad \text{at}\quad z=\pm\tfrac12. 
\label{eq: no slip}
\eeq
At a boundary where $\vu$ satisfies the fixed-stress condition~\cref{eq: stress free u}, the $\vvv$ condition is stress-free,
\beq
\partial_z v_1=\partial_z v_2=v_3 = 0 \quad \text{at}\quad z=\pm\tfrac12. 
\label{eq: stress free}
\end{equation}
Formulations of the background method in the present section assume that~\cref{eq: no slip} or~\cref{eq: stress free} holds at each boundary, but the two boundaries need not be the same. Results in \cref{sec: main argument} require the same condition at both boundaries.

\subsection{Background method formulation in terms of auxiliary functionals}
\label{sec: background method}

For shear flows governed by \cref{eq: NSE}, the quantity that has most often been bounded using the background method is the mean dissipation. We let angle brackets denote an average over the spatial domain $\Omega$ and let an overbar denote an infinite-time average, so the dimensionless dissipation averaged over the volume and infinite time is
\begin{equation}
	\overline{\left\langle|\nabla\vu|^2\right\rangle}=\lim_{T\to\infty}\frac{1}{T}\int_0^T\frac{1}{\Gamma_x\Gamma_y}\int_{\Omega}|\nabla\vu|^2\,\dx\,\dt,
	\label{eq: mean diss}
\end{equation}
where $\Gamma_x\Gamma_y$ is the volume of the dimensionless domain and, if the infinite-time limit is not well-defined one can take the limit supremum. The dimensionless quantity \cref{eq: mean diss} is the same one used in previous works such as that of \citet{plasting_improved_2003}. For physical interpretation in \cref{sec: computational}, we examine this dissipation's ratio to its laminar value since this ratio does not depend on the chosen nondimensionalization. The average dissipation is related \emph{a priori} to the average rate of energy input by body and boundary forces, as explained for the examples of Wallefe flow and Couette flow in \cref{sec: computational}. 

Whether one seeks upper or lower bounds depends on the model. Often it is easy to show that the dissipation among all flows is bounded below or above by the laminar dissipation, in which case it remains only to find upper or lower bounds, respectively. For concreteness, our exposition in this section considers upper bounds on mean dissipation. \Cref{sec: lower bounds} summarizes what changes in the case of lower bounds. The background method can bound averages of other linear or quadratic integrals also, such as kinetic energy, by straightforward modifications to the formulations given here.

Our goal is to derive bounds on \cref{eq: mean diss} that apply to all solutions of the governing equations \cref{eq: NSE} subject to boundary conditions, regardless of the initial conditions. We give the derivation in terms of a so-called auxiliary functional $V$. Such functionals have not been explicitly used in the background method literature until recently, but they are implicit in all such arguments, as explained by \citet{chernyshenko_relationship_2022} and \citet{fantuzzi_background_2021}. An auxiliary functional $V[\vw]$ maps a divergence-free, time-independent vector field $\vw(\vx)$ to a real number. All past applications of the background method to planar shear flows are equivalent to choosing $V[\vw]$ that is a quadratic spatial integral of the form
\begin{equation}
	V[\vw]=\Rey\left\langle\tfrac{a}{2}|\vw|^2- \zeta\hat{\vx}\cdot\vw\right\rangle,
	\label{eq: V}
\end{equation}
where the functional is defined by choosing the `balance parameter' $a$ and the `background field' $\zeta(z)\hat{\vx}$. The quantity $\tfrac{1}{a}\zeta$ is called the background field in the notation of most prior works \citep[cf.][]{chernyshenko_relationship_2022}, but it is useful for what follows to define notation for $\zeta$ rather than for $\tfrac{1}{a}\zeta$ because then, $a$ and $\zeta$ appear linearly in $V$. Generalizations of \cref{eq: V} that go beyond quadratic integrals have the potential to give stronger results, as they do for the Kuramoto--Sivashinsky equation \citep{goluskin2019bounds}, but the present work concerns the background method and thus only $V$ of the form \cref{eq: V}. There is no advantage to considering a background field of more general form than $\zeta(z)\hat{\vx}$, as proved in \cref{app: zeta 1D} using symmetry arguments. To enable integration by parts later, the background field is admissible only if it is continuous and piecewise smooth, and if it satisfies the same boundary conditions as $\vvv(\vx,t)$.

The definition of $V$ does not involve time, but one obtains a scalar-valued function of time by considering $V[\vvv(\vx,t)]$, where $\vvv$ solves \cref{eq: NSE v}. For choices of $V$ that lead to finite bounds on mean dissipation, it can be shown that $V[\vvv(\vx,t)]$ remains bounded as $t\to\infty$ for any admissible initial condition $\vvv(\vx,0)$ \citep{doering_variational_1994}. All such $V$ satisfy the identity
\beq
\overline{\tfrac{\rm d}{\dt}V[\vvv(\vx,t)]}=\lim_{T\to\infty}\frac{1}{T}\left(V[\vvv(\vx,T)]-V[\vvv(\vx,0)]\right)=0,
\label{eq: V identity}
\eeq
so the time-averaged dissipation $\overline{\langle |\nabla\vu|^2\rangle}$ that we want to bound is equal to $\overline{\langle |\nabla\vu|^2\rangle+\tfrac{\rm d}{\dt}V}$. Moreover, one can find an expression equal to $\tfrac{\rm d}{\dt}V[\vvv]$ without explicit time-dependence:
\begin{align}
	\tfrac{\rm d}{\dt}V[\vvv(\vx,t)] 
	&= \Rey\left\langle (a\vvv-\zeta\vxhat)\cdot \partial_t\vvv\right\rangle \label{eq: first line} \\
	&=\Rey\left\langle (a\vvv-\zeta\vxhat)\cdot \left(- \vvv\cdot\nabla\vvv - U\partial_x\vvv - U'v_3\hat{\vx} -\nabla p + \tfrac{1}{\Rey}\lap\vvv \right)\right\rangle \label{eq: second line} \\
	&= \left\langle-a|\nabla\vvv|^2+\zeta'\partial_zv_1-\Rey\left(aU+\zeta\right)'v_1v_3\right\rangle. \label{eq: dVdt 3}
\end{align}
Equation \eqref{eq: first line}  is derived by moving the time derivative inside the volume integral, and \eqref{eq: second line} replaces $\partial_t\vvv$ according to \cref{eq: NSE v}. Equation \eqref{eq: dVdt 3} follows from integration by parts, recalling that we require $\zeta(z)\vxhat$ to satisfy the same boundary conditions as~$\vvv$, which may be stress-free or no-slip at each boundary. Finally, we find a useful expression with the same time average as the dissipation:
\begin{align}
	\overline{\langle|\nabla\vu|^2\rangle}
	&=\overline{\langle U'^2+2U'\partial_zv_1+|\nabla\vvv|^2\rangle} \label{eq: diss identity 1} \\
	&= \overline{\langle U'^2+2U'\partial_zv_1+|\nabla\vvv|^2\rangle+\tfrac{d}{dt}V[\vvv(\vx,t)]} \label{eq: diss identity 2} \\
	&= \overline{\funcq[\vvv(\vx,t)]}, \label{eq: diss identity 3}
\end{align}
where the functional $\funcq$ is defined as
\begin{equation}
\funcq[\vw]=\big\langle U'^2+(2U+\zeta)'\partial_zw_1-(a-1)|\nabla\vw|^2-\Rey(aU+\zeta)'w_1w_3\big\rangle
\label{eq: Q}
\end{equation}
for any time-independent, divergence-free vector field $\vw=(w_1,w_2,w_3)$ that obeys the same boundary conditions as $\vvv$. The equality \cref{eq: diss identity 1} follows from substituting $\vu=U\vxhat+\vvv$, the next equality uses \cref{eq: V identity}, and then applying \cref{eq: dVdt 3} gives the expression for $\funcq[\vw]$.

Since the time average $\overline{\funcq[\vvv]}$ is bounded above by the maximum of $\funcq[\vw]$ over all possible~$\vw$, equation~\cref{eq: diss identity 3} implies
\begin{equation}
\overline{\langle|\nabla\vu|^2\rangle} \le \max_{\vw\in\mathcal{H}_{3D}}\funcq[\vw]
\label{eq: UB}
\end{equation}
for any $a$ and any admissible $\zeta$, where $\mathcal{H}_{3D}$ is the class of \mbox{3-D} divergence-free vector fields $\vw(\vx)$ that satisfy the same boundary conditions~\cref{eq: no slip} or~\cref{eq: stress free} as $\vvv(\vx,t)$ does. The right-hand maximum in~\cref{eq: UB} can be finite or infinite, depending on $a$ and $\zeta$. When the maximum is finite, it can be computed numerically or bounded above analytically. One naturally wants to choose $a$ and $\zeta$ to make the resulting upper bound as small as possible. At each fixed $\Rey$, the upper bound on dissipation that is optimal within the framework of the background method is the solution to a min--max problem,
\begin{equation}
\overline{\left\langle|\nabla\vu|^2\right\rangle}\leq\min_{\substack{a,\,\zeta}}\max_{\vw\in\mathcal{H}_{3D}}\funcq[\vw],
\label{eq: UB opt}
\end{equation}
where the function class over which $\zeta$ is minimized is such that $\zeta\vxhat\in\mathcal{H}_{3D}$. In \cref{sec: decoupling of mean flow,sec: spectral constraint} we give three more formulations that are equivalent to~\cref{eq: UB opt} and are useful for different purposes.

The optimal background method formulation \cref{eq: UB opt} is a more precise version of \cref{eq: intro} for the case of planar shear flows. Since the inner maximization gives an infinite value for some $a$ and $\zeta$, strictly speaking the minimum and maximum should be called an infimum and supremum, respectively, but we use the notation $\min$ and $\max$ throughout. Provided these values are finite, we assume that all maxima and minima are attained, meaning that there exist $a$, $\zeta$ and $\vw$ for which $\funcq[\vw]$ is equal to its min--max in~\cref{eq: UB opt}. Such attainment relies on optimizing $\zeta$ and $\vw$ over sufficiently large function spaces \citep{evans2022partial}, but the exact choice of these spaces is beyond our scope.

\subsection{Mean and mean-free decomposition of $\funcq$}
\label{sec: decoupling of mean flow}

The maximization over $\vw$ in~\cref{eq: UB} can be decoupled into maximizations over the planar mean of $\vw$ and over its remaining mean-free part. This decoupling leads to a variational problem familiar from energy stability analysis, eventually allowing us to apply the theorem of \citet{busse_property_1972}. It also reveals that finite values of the upper bound~\cref{eq: UB} depend on $a$ and $\zeta$ but not on $\Rey$, even though the functional $\funcq$ being maximized depends also on $\Rey$. For fixed $a$ and $\zeta$, the maximum will assume a constant value at sufficiently small $\Rey$, and for larger $\Rey$ the right-hand side of~\cref{eq: UB} will be infinite. 

We decompose the divergence-free vector field $\vw$ as 
\begin{equation}
	\vw(\vx)=\vF(z)+\dot{\vw}(\vx),
	\label{eq: w decomp}
\end{equation}
where $\vF$ is the mean of $\vw$ over the periodic $x$ and $y$ directions, and $\dot\vw$ is the remaining mean-free part. We denote the components of the mean as $\vF=(F_1,F_2,0)$, where the zero wall-normal component follows from impenetrability of the walls and incompressibility. With $\vw$ so decomposed, the $\funcq$ functional defined by~\cref{eq: Q} decouples into functionals of $\vF$ and of $\dotw$,
\begin{equation}
\label{eq: Q expressed mean and mean free}
\funcq[\vF+\dot\vw]=\left\langle U'^2\right\rangle+\funcf[\vF]+\funce[\dotw],
\end{equation}
where
\begin{align}
\funcf[\vF] &= \left\langle(2U+\zeta)'F'_1-(a-1)\big({F'_1}^2+{F'_2}^2\big)\right\rangle, \label{eq: F} \\
\funce[\dot\vw] &= \left\langle -(a-1)|\nabla\dot\vw|^2-\Rey\,(aU+\zeta)'\dot{w}_1\dot{w}_3\right\rangle. \label{eq: E}
\end{align}
To find the maximum of $\funcq[\vw]$, we can separately maximize $\funcf[\vF]$ and $\funce[\dotw]$.

The functional $\funcf[\vF]$ has a finite maximum only if $a>1$, so we assume this hereafter. Since the $F_2$ term in~\cref{eq: F} is non-positive and independent of other terms, the maximizer $\vF^*$ that attains the maximum of $\funcf$ must have $F{^*_2}'=0$. With no-slip boundaries this implies $F^*_2=0$, and with stress-free boundaries we let $F^*_2=0$ to fix the reference frame. The Euler--Lagrange equation for $F_1$ then gives $F_1^*=\frac{1}{2(a-1)}\left(2U+\zeta\right)$. Thus,
\begin{equation}
\label{eq:maximization over mean free part}
\max_{\vw\in\mathcal{H}_{3D}}\funcq[\vw] = 
\max_{\dot\vw\in\dot{\mathcal{H}}_{3D}}\funcq[\vF^*+\dot\vw]=
\tfrac{1}{a-1}\big\langle a{U'}^2+U'\zeta'+\tfrac14{\zeta'}^2\big\rangle + 
\max_{\dotw\in\dot{\mathcal{H}}_{3D}}\funce[\dotw],
\end{equation}
where $\dot{\mathcal{H}}_{3D}$ is the mean-free subspace of $\mathcal{H}_{3D}$. Henceforth we do not write $\dotw$; fields denoted by $\vw$ may be mean-free or not, depending on their function spaces.

The maximum in~\cref{eq:maximization over mean free part} is finite---thus giving an upper bound on dissipation---if and only if the $\funce$ functional is non-positive for all mean-free fields. If $\funce$ is non-positive, its maximum of zero is attained by $\vw=\mathbf 0$. Otherwise its maximum is infinity since all terms in $\funce$ are quadratic; any $\vw$ for which $\funce[\vw]>0$ can be scaled to make $\funce$ arbitrarily large. With this observation, minimizing~\cref{eq:maximization over mean free part} over $a$ and $\zeta$ gives our second formulation of the optimal background method bound,
\begin{equation}
\overline{\left\langle|\nabla\vu|^2\right\rangle}\leq\min_{\substack{a>1 \\ \zeta(z)}}
~\tfrac{1}{a-1}\big\langle a{U'}^2+U'\zeta'+\tfrac14{\zeta'}^2\big\rangle \quad \text{s.t.}\quad 
\funce[\vw]\le0 ~~\forall\,\vw\in\dot{\mathcal{H}}_{3D}.
\label{eq: UB opt spectral}
\end{equation}
The constrained minimization in the second formulation~\cref{eq: UB opt spectral} is equivalent to the unconstrained min--max in the first formulation~\cref{eq: UB opt}. The $\funce\le0$ constraint in~\cref{eq: UB opt spectral} is called the spectral constraint because, as shown in the next subsection, it can be formulated in terms of the spectrum of a linear eigenproblem.

\subsection{Spectral constraint}
\label{sec: spectral constraint}

We now derive two more ways to formulate the spectral constraint in~\cref{eq: UB opt spectral}, thus giving a third and fourth formulation of the optimal background method problem. Both reformulations are familiar from energy stability analysis of shear flows. To see the connection to energy stability, note that non-positivity of the $\funce$ functional defined by~\cref{eq: E} is equivalent to
\beq
-\left\langle |\nabla\vw|^2\right\rangle - \Rey \left\langle hw_1w_3\right\rangle \le0 ~~\forall\,\vw\in\dot{\mathcal{H}}_{3D},
\label{eq: spectral h}
\eeq
where we have divided by the positive quantity $a-1$ and let
\beq
h(z) = \tfrac{1}{a-1}\left[aU'(z)+\zeta'(z)\right].
\label{eq: h}
\eeq
The condition~\cref{eq: spectral h} is precisely the energy stability condition for a laminar flow whose derivative is $h(z)$ rather than $U'(z)$. For such a flow, the time derivative of the perturbation energy $\left\langle|\vvv|^2\right\rangle$ is non-positive for all admissible $\vvv(\vx,t)$ if and only if~\cref{eq: spectral h} holds. In other words, the spectral constraint of the background method for a flow with laminar shear $U'$ is exactly the energy stability constraint for a flow with laminar shear $h$. Just as laminar flows are energy stable only below a certain $\Rey$ value that depends on $U$, the spectral constraint is satisfied only below a certain $\Rey$ that depends on $U$, $a$ and $\zeta$.

One way to reformulate the spectral constraint is to rearrange~\cref{eq: spectral h} as an inequality for $\Rey$. Note that the first term of~\cref{eq: spectral h} is negative definite, while the second is sign-indefinite. In fact, for any field $[w_1,w_2,w_3](x,y,z)$ giving certain values for first and second terms in~\cref{eq: spectral h}, the field $[-w_1,w_2,w_3](-x,y,z)$ gives the same first term but negates the second. Therefore, taking the absolute value of the second term in~\cref{eq: spectral h} gives an equivalent condition,
\beq
-\left\langle |\nabla\vw|^2\right\rangle + \Rey \left|\left\langle hw_1w_3\right\rangle\right| \le0 ~~\forall\,\vw\in\dot{\mathcal{H}}_{3D}.
\label{eq: spectral abs}
\eeq
In turn, this is equivalent to
\beq
\Rey \le \min_{\vw\in\dot{\mathcal{H}}_{3D}}\funcr[\vw],
\label{eq: spectral Re}
\eeq
where
\beq
\funcr[\vw]=\frac{\left\langle |\nabla\vw|^2\right\rangle}{\left|\left\langle hw_1w_3\right\rangle\right|}.
\label{eq: R}
\eeq

Expressing the spectral constraint in~\cref{eq: UB opt spectral} using $\funcr$ gives our third formulation of the optimal background method,
\begin{equation}
\overline{\left\langle|\nabla\vu|^2\right\rangle}\leq\min_{\substack{a>1 \\ \zeta(z)}}
~\tfrac{1}{a-1}\big\langle a{U'}^2+U'\zeta'+\tfrac14{\zeta'}^2\big\rangle \quad \text{s.t.}\quad 
\Rey\le \min_{\vw\in\dot{\mathcal{H}}_{3D}}\funcr[\vw].
\label{eq: UB opt spectral R}
\end{equation}
For fixed $a$ and $\zeta$, the background method gives the right-hand integral in~\cref{eq: UB opt spectral} as an upper bound if $\Rey$ satisfies~\cref{eq: spectral Re}, and it gives no finite bound at larger $\Rey$. The energy stability criterion is commonly expressed as~\cref{eq: spectral Re} for a flow with laminar shear profile $h$. This is the formulation of energy stability used throughout the analysis of \citet{busse_property_1972}, which underlies our criterion derived in \cref{sec: main argument}. We thus express the spectral constraint as~\cref{eq: spectral Re} throughout \cref{sec: main argument}.

Another reformulation of the spectral constraint is in terms of the spectrum of a linear eigenproblem. Since $\funce[\vw]\le0$ holds if and only if it holds when $\vw$ is scaled to satisfy $\left\langle|\vw|^2\right\rangle=1$, the spectral constraint is equivalent to
\beq
\min_{\substack{\vw\in\dot{\funch}_{3D}\\\langle|\vw|^2\rangle=1}}\left(-\funce[\vw]\right)\ge0.
\label{eq: spectral normalized}
\eeq
We have negated $\funce$ to obtain a non-negativity condition for consistency with prior works. The normalization constraint on $\vw$ has been added so that the left-hand minimum is negative but still finite when the spectral constraint is violated. Then, whether or not the spectral constraint holds, the variational problem~\cref{eq: spectral normalized} has mean-free minimizers that satisfy its Euler--Lagrange equations,
\begin{equation}
\label{eq: Spectral constraint EVP}
-(a-1)\lap\vw+\tfrac12\Rey(aU+\zeta)'[w_3\hat\vx+w_1\hat{\vz}]+\nabla p=\lambda\vw,
\quad \nabla\cdot\vw=0, 
\end{equation}
where $2p(\vx)$ is a Lagrange multiplier enforcing incompressibility, and $\lambda$ is a Lagrange multiplier enforcing $\langle|\vw|^2\rangle=1$ that acts as an eigenvalue in~\cref{eq: Spectral constraint EVP}. The spectral constraint is equivalent to all eigenvalues of~\cref{eq: Spectral constraint EVP} being non-negative---that is, to this linear eigenproblem having a non-negative spectrum.

To simplify implementation of the eigenproblem~\cref{eq: Spectral constraint EVP}, one can Fourier transform in the periodic directions. This gives an eigenproblem that is an ordinary differential equation in $z$ for each fixed wavevector $\wn=(j,k)$, where $j$ and $k$ are the streamwise and spanwise wavenumbers, respectively. For each $\wn$,
\begin{equation}
\label{eq: fourier transformed eigenvalue problem}
\begin{split}
-(a-1)\left(\tfrac{\rm d^2}{{\rm d}z^2}-j^2-k^2\right) \hat{w}_1+\tfrac12\Rey (aU+\zeta)'\hat w_3+ij\hat{p}&=\lambda \hat{w}_1, \\ 
-(a-1)\left(\tfrac{\rm d^2}{{\rm d}z^2}-j^2-k^2\right)  \hat{w}_2+ik\hat{p}&=\lambda \hat{w}_2, \\ 
-(a-1)\left(\tfrac{\rm d^2}{{\rm d}z^2}-j^2-k^2\right)  \hat{w}_3+\tfrac12\Rey (aU+\zeta)'\hat w_1+\tfrac{\rm d}{{\rm d}z}\hat{p}&=\lambda \hat{w}_3, \\
		ij\hat{w}_1+ik\hat{w}_2+\tfrac{\rm d}{{\rm d}z}\hat{w}_3&=0,
	\end{split}
\end{equation}
where $i$ is the imaginary unit, the Fourier transforms $\hat{\vw}(z)$ and $\hat{p}(z)$ are complex in general, all $\lambda$ are real,  and $\tfrac{\rm d}{{\rm d}z}$ is the ordinary $z$-derivative operator. The spectral constraint is equivalent to~\cref{eq: fourier transformed eigenvalue problem} having a non-negative spectrum of $\lambda$ for all admissible $\wn$. This gives our fourth formulation of the optimal background method,
\begin{equation}
\overline{\left\langle|\nabla\vu|^2\right\rangle}\leq\min_{\substack{a>1 \\ \zeta(z)}}
~\tfrac{1}{a-1}\big\langle a{U'}^2+U'\zeta'+\tfrac14{\zeta'}^2\big\rangle
\quad \text{s.t.}\quad 
\lambda\ge0 \text{ in }\cref{eq: fourier transformed eigenvalue problem}~\forall\,\wn\in K,
\label{eq: UB opt spectral eigen}
\end{equation}
where $K$ is the set of admissible wavevectors $\wn=(j,k)$. It suffices for $K$ to include only non-negative $j$ and $k$; adding constraints for $(-j,k)$, $(j,-k)$ or $(-j,-k)$ would be redundant. The velocity is mean-free in $x$ and $y$, so $(0,0)\notin K$. For bounds to apply for all possible periods $\Gamma_x$ and $\Gamma_y$, the spectrum of~\cref{eq: UB opt spectral eigen} must be non-negative for all other non-negative pairs $(j,k)$. For bounds to apply to flows with fixed $\Gamma_x$ and $\Gamma_y$, the spectrum of~\cref{eq: UB opt spectral eigen} must be non-negative only for $j$ and $k$ that are integer multiples of $2\pi/\Gamma_x$ and $2\pi/\Gamma_y$. Enforcing the spectral constraint for only \mbox{2.5-D} fields amounts to including only $(0,k)\in K$; this is what \citet{plasting_improved_2003} did when computing optimal bounds for Couette flow.

The main question of our present work---whether bounds computed over \mbox{2.5-D} and \mbox{3-D} fields coincide---will have the same answer for all four equivalent formulations of the optimal background method in \cref{eq: UB opt,eq: UB opt spectral,eq: UB opt spectral R,eq: UB opt spectral eigen}. The same is true for suboptimal bounds found by maximizing over $\vw$ at fixed $a$ and $\zeta$. Our computations in \cref{sec: computational} implement the fourth formulation~\cref{eq: UB opt spectral eigen} to find the optimal $a$ and $\zeta$ under the assumption of \mbox{2.5-D} optimizers, meaning all wavevectors in $K$ have the form $(0,k)$. One can confirm \emph{a posteriori} that these bounds apply to \mbox{3-D} flows by directly checking that the spectral constraint in~\cref{eq: UB opt spectral eigen} holds also for $(j,k)$ with $j\neq0$. In some cases, however, direct checking of the \mbox{3-D} spectral constraint can be avoided by using the criterion derived in \cref{sec: main argument}.

\subsection{Lower bounds}
\label{sec: lower bounds}

The four formulations of the background method derived in \cref{sec: background method,sec: decoupling of mean flow,sec: spectral constraint} require only slight modification to give lower bounds on dissipation rather that upper bounds. In the first formulation~\cref{eq: UB opt}, we simply switch the role of minimization and maximization to find
\begin{equation}
\overline{\left\langle|\nabla\vu|^2\right\rangle}\geq\max_{\substack{a,\,\zeta}}\min_{\vw\in\mathcal{H}_{3D}}\funcq[\vw].
\label{eq: LB opt}
\end{equation}
Decomposition of $\vw$ into its mean and mean-free parts as in \cref{sec: decoupling of mean flow} leads to the lower bound version of the second formulation
\begin{equation}
\overline{\left\langle|\nabla\vu|^2\right\rangle}\geq\max_{\substack{a<1 \\ \zeta(z)}}
~\tfrac{1}{a-1}\big\langle a{U'}^2+U'\zeta'+\tfrac14{\zeta'}^2\big\rangle \quad \text{s.t.}\quad 
\funce[\vw]\ge0 ~~\forall\,\vw\in\dot{\mathcal{H}}_{3D}.
\label{eq: LB opt spectral}
\end{equation}
This differs from its analogue~\cref{eq: UB opt spectral} for upper bounds only in that $\funce$ must have the opposite sign, and finite bounds require $a<1$ rather than $a>1$. Dividing the expression~\cref{eq: E} for $\funce$ by the positive quantity $1-a$, as opposed to $a-1$ in the upper bound case, and giving the second term its worst-case sign as in \cref{sec: spectral constraint}, we find that non-negativity of $\funce[\vw]$ is equivalent to
\beq
\left\langle |\nabla\vw|^2\right\rangle - \Rey \left|\left\langle hw_1w_3\right\rangle\right| \ge0 ~~\forall\,\vw\in\dot{\mathcal{H}}_{3D},
\label{eq: spectral abs LB}
\eeq
where $h(z)$ is defined by~\cref{eq: h} as for upper bounds. Rearranging~\cref{eq: spectral abs LB} as an upper bound on $\Rey$, we find that the spectral constraint is identical in the upper and lower bound cases, so the lower bound version of the third formulation is
\begin{equation}
\overline{\left\langle|\nabla\vu|^2\right\rangle}\geq\max_{\substack{a<1 \\ \zeta(z)}}
~\tfrac{1}{a-1}\big\langle a{U'}^2+U'\zeta'+\tfrac14{\zeta'}^2\big\rangle \quad \text{s.t.}\quad 
\Rey\le \min_{\vw\in\dot{\mathcal{H}}_{3D}}\funcr[\vw].
\label{eq: LB opt spectral R}
\end{equation}
The fourth formulation is the same as~\cref{eq: LB opt spectral R}, except the spectral constraint takes its eigenvalue form as in~\cref{eq: UB opt spectral eigen}. In \cref{sec: main argument}, theoretical results apply to both upper and lower bounds because the spectral constraint takes the same form in both cases.

\section{A criterion for streamwise invariance of optimizers}
\label{sec: main argument}

This section presents our main theoretical result: a criterion for confirming that optima over \mbox{2.5-D} fields and over \mbox{3-D} fields coincide in the background method for shear flow models with a certain symmetry. This symmetry is defined in \cref{sec: sym}, then \cref{sec: busse result} explains how our criterion follows from a criterion of Busse concerning energy stability eigenproblems. \Cref{sec: primitive} summarizes a computational procedure where optimal bounds are computed over \mbox{2.5-D} fields, some additional easier computations are carried out, and then our criterion is used to verify that the bounds hold for \mbox{3-D} flows. \Cref{sec: restricted background profiles} describes a different approach, where our criterion is included as a constraint in the original \mbox{2.5-D} bounding computations.

\subsection{Assumed symmetry of the governing equations}
\label{sec: sym}

Throughout \cref{sec: main argument} we assume that the governing model is invariant under $180^{\circ}$ rotation about a spanwise axis,
\begin{equation}
[u_1,u_2,u_3](x,y,z,t)\mapsto[-u_1,u_2,-u_3](-x,y,-z,t).
\label{eq: sym}
\end{equation}
That is, we assume that the left-hand side of \cref{eq: sym} satisfies the governing equation~\cref{eq: NSE} and boundary conditions if and only if the right-hand side satisfies them. This requires that any body forcing $f(z)$ must be odd about the $z=0$ mid-plane. It also requires the boundary conditions to be odd, meaning that if $u_1$ values are fixed at $z=\pm1/2$ then they must be negations of each other, and likewise if $\partial_zu_1$ values are fixed. The flow itself need not be invariant under~\cref{eq: sym}.

For all shear flow models invariant under~\cref{eq: sym}, the laminar flow profile $U(z)$ is odd about the mid-plane, and the equations~\cref{eq: NSE v} and the boundary conditions governing perturbations $\vvv(\vx,t)$ are also invariant under~\cref{eq: sym}. For shear flows with this symmetry we always restrict to background profiles $\zeta(z)$ that are odd because this cannot worsen the eventual upper bound, as shown in \cref{app: zeta odd}.

\subsection{Busse's criterion}
\label{sec: busse result}

For our present purpose we consider the third formulation of the optimal background method bound, where the spectral constraint requires that $\Rey$ is no larger than the minimum of $\funcr[\vw]$. This constraint is identical in the upper and lower bounding formulations of~\cref{eq: UB opt spectral R,eq: LB opt spectral R}. If this minimum is the same over \mbox{2.5-D} and \mbox{3-D} fields for given $\zeta$ and $a$, then in any formulation of the background method it suffices to consider \mbox{2.5-D} fields. In particular, we aim to compute the optimal $\zeta$ and $a$ using \mbox{2.5-D} fields and then verify that the resulting bounds are valid also for \mbox{3-D} fields. 

The minimum of $\funcr[\vw]$ in~\cref{eq: UB opt spectral R} is exactly the critical $\Rey$ value of energy stability for a flow whose laminar profile is $\tfrac{1}{a-1}(aU+\zeta)$ rather than $U$. This follows from the relationship between the spectral constraint and energy stability that is described in the first paragraph of~\cref{sec: spectral constraint} above. Thus our present question about the background method amounts to asking whether the critical $\Rey$ of energy stability for the laminar profile$\tfrac{1}{a-1}(aU+\zeta)$ is the minimum of $\funcr[\vw]$ over \mbox{2.5-D} fields, or whether \mbox{3-D} fields would give a smaller minimum. \Citet{busse_property_1972} derived a criterion for drawing exactly this conclusion about the energy stability problem for models with the symmetry~\cref{eq: sym}. That criterion is directly applicable here since we are interested in energy stability of the laminar profile $\tfrac{1}{a-1}(aU+\zeta)$, which has the required symmetry. In particular, this profile is odd, and its derivative $h$ is even, because the symmetry~\cref{eq: sym} ensures that $U$ is odd and lets us restrict to odd $\zeta$.

The criterion of \citeauthor{busse_property_1972} is most naturally stated using poloidal--toroidal variables, on which its derivation relies. Any divergence-free and mean-free $\vw\in\dot{\mathcal{H}}_{3D}$ in the present geometry can be represented as
\begin{equation}
	\vw=\ccurl(\varphi\vzhat)+\curl(\psi\vzhat),
	\label{eq: polo tor}
\end{equation}
where $\varphi(\vx)$ and $\psi(\vx)$ are the poloidal and toroidal potentials, respectively. These potentials are uniquely determined by $\vw$, up to arbitrary additive constants. See \citet{heywood_decomposition_1992} for a proof that this decomposition is always possible with the present geometry and boundary conditions. In these variables, no-slip conditions~\cref{eq: no slip} at both boundaries are
\begin{equation}
	\label{potential  function boundary conditions}
	\varphi,\:\partial_z\varphi,\:\psi=0\quad\text{at}\quad z=\pm\tfrac{1}{2},
\end{equation}
and stress-free conditions~\cref{eq: stress free} at both boundaries are
\begin{equation}
	\label{potential  function boundary conditions stress}
	\varphi,\:\partial_z^2\varphi,\:\partial_z\psi=0\quad\text{at}\quad z=\pm\tfrac{1}{2}. 
\end{equation}
In terms of $\varphi$ and $\psi$, the $\funcr$ functional defined by~\cref{eq: R} can be expressed as
\begin{equation}
	\label{eq: R pol tor}	
	\funcr[\vw]=\frac{\left\langle|\vzhat\times\nabla\lap\varphi|^2+|\ccurl(\psi\vzhat)|^2\right\rangle}{\left|\left\langle h\,\lap_2\varphi\left(\partial_y\psi+\partial_{xz}\varphi\right)\right\rangle\right|},
\end{equation}
where $\lap_2=\partial_x^2+\partial_y^2$ denotes the Laplacian operator in only the periodic directions. The derivation of~\cref{eq: R pol tor} from~\cref{eq: R} has used the identity $\langle|\nabla\vw|^2\rangle=\langle|\curl\vw|^2\rangle$ that holds for both no-slip and stress-free boundaries.

The criterion of \citet{busse_property_1972} gives a sufficient condition for the $\funcr[\vw]$ functional defined by~\cref{eq: R}, or by~\cref{eq: R pol tor} in poloidal--toroidal variables, to have the same minimum over \mbox{2.5-D} and \mbox{3-D} fields. It applies in the present geometry if $h(z)$ is an even function, and if the minimizations over $\vw$ allow for all spanwise or streamwise periods. The condition can be checked by computing minima of $\funcr[\vw]$ over three different lower-dimensional subspaces to find the following values:
\begin{align}
	R_e=\min_{\substack{\vw\in\Hdot_{2.5D}\\\varphi,\psi\text{ even in }z}}\funcr[\vw], \quad~
	R_o=\min_{\substack{\vw\in\Hdot_{2.5D}\\\varphi,\psi\text{ odd in }z}}\funcr[\vw], \quad~
	R_\varphi=\min_{\substack{\vw\in\Hdot_{2D}\\\psi=0}}\funcr[\vw].
	\label{eq: 3 mins}
\end{align}
The theorem of \citet{busse_property_1972} can be stated as:
\beq
\text{If}\quad \frac{1}{R_e^2}\ge\frac{1}{R_o^2}+\frac{1}{R_\varphi^2}, 
\quad\text{then}\quad \min_{\vw\in\Hdot_{3D}}\funcr[\vw] = R_e.
\label{eq: busse criterion}
\eeq
The proof of this statement is agnostic to whether $h=U'$, as in Busse's context of energy stability, or $h= \tfrac{1}{a-1}[aU'+\zeta']$, as in the spectral constraint of the background method. Observe that $R_e$ and $R_o$ are computed over different subspaces of \mbox{2.5-D} fields $\vw(y,z)$, and $R_\varphi$ is computed over a subspace of \mbox{2-D} fields $\vw(x,z)$. It is explained in \cref{app: euler lagrange} why Busse's proof of~\cref{eq: busse criterion} requires that the minimizations over \mbox{2.5-D} and \mbox{2-D} fields admit all spanwise and streamwise periods, respectively. The inequality in~\cref{eq: busse criterion} is what we refer to as `Busse's criterion'. Later, we use an equivalent form of Busse's criterion in which the inequality is rearranged as
\beq
\chi\le 1, \quad\text{where}\quad
\chi = R_e \left(\tfrac{1}{R_o^2}+\tfrac{1}{R_\varphi^2}\right)^{1/2}.
\label{eq: chi}
\eeq

A proof of~\cref{eq: busse criterion} is given in \cref{app: proof of busse theorem}, where we follow the same approach as \citet{busse_property_1972} with some details added or changed for clarity. At present we explain only the last part of the argument. The desired equality in~\cref{eq: busse criterion} certainly holds as an inequality,
\beq
\min_{\vw\in\Hdot_{3D}}\funcr[\vw]
\le \min_{\substack{\vw\in\Hdot_{2.5D}\\\varphi,\psi\text{ even in }z}} \funcr[\vw]\equiv R_e,
\label{eq: R_e geq 3D}
\eeq
since the space of $\vw$ for the right-hand minimization is contained in the space for the left-hand minimization. With a much longer argument, it is shown in \cref{app: proof of busse theorem} that the minimum of interest over \mbox{3-D} fields is bounded below by
\begin{equation}
	\min_{\vw\in\Hdot_{3D}}\funcr[\vw]\geq 
	\min\left\{R_e,\sqrt\beta R_o,\sqrt{1-\beta}R_\varphi\right\}
	\label{eq: pre main result}
\end{equation}
for every $\beta\in(0,1)$. The inequality opposite to~\cref{eq: R_e geq 3D} holds, thereby giving equality, if $R_e$ is the smallest of the three values on the right-hand side of~\cref{eq: pre main result}. There exists a choice of $\beta\in(0,1)$ for which $R_e$ is the smallest value, meaning $R_e\le\sqrt{\beta}R_o$ and $R_e\le\sqrt{1-\beta}R_\varphi$, if and only if the inequality in~\cref{eq: busse criterion} holds. (One can choose $\beta=R_e^2/R_o^2$.) Therefore, Busse's criterion~\cref{eq: busse criterion} indeed follows once the inequality~\cref{eq: busse criterion} is proved in \cref{app: proof of busse theorem}.

The version of Busse's theorem formulated in~\cref{eq: busse criterion} aims to conclude that the minimum of $\funcr$ over \mbox{3-D} fields is attained by \mbox{2.5-D} fields whose poloidal and toroidal potentials are even in~$z$. This is indeed the case for various $h(z)$ arising in energy stability analysis, and for some of the $h(z)$ arising in our background method computations of \cref{sec: computational}. A criterion for the minimum to be attained by \mbox{2.5-D} fields whose potentials are instead odd in~$z$ can be derived by switching the roles of $R_e$ and $R_o$ in the proof of \cref{app: proof of busse theorem}. This gives the following:
\beq
\text{If}\quad \frac{1}{R_o^2}\ge\frac{1}{R_e^2}+\frac{1}{R_\varphi^2}, 
\quad\text{then}\quad \min_{\vw\in\Hdot_{3D}}\funcr[\vw] = R_o.
\label{eq: busse criterion 2}
\eeq
The only difference between~\cref{eq: busse criterion 2} and~\cref{eq: busse criterion} is swapping oddness and evenness. We are not aware of examples where the minimum of $\funcr$ over \mbox{3-D} fields coincides with $R_o$ rather than $R_e$, but we have not ruled them out.

\subsection{Application of Busse's criterion to the background method}
\label{sec: primitive}

We now describe how the criterion in \cref{sec: busse result} can be used to ensure that \mbox{2.5-D} background method computations apply to \mbox{3-D} flows for any model with the symmetry~\cref{eq: sym}. A calculation over \mbox{2.5-D} fields at chosen $\Rey$ gives $a$ and $\zeta$. For this combination of $\Rey$, $a$ and $\zeta$, one must confirm that the extremum of $\funcq$ is the same over \mbox{3-D} fields as over \mbox{2.5-D} fields. In both the upper and lower bounding cases, this is equivalent to $\funcr$ having the same minimum over \mbox{3-D} and \mbox{2.5-D} fields. The latter can be verified using the criterion of \cref{sec: busse result} as part of the following procedure.\\
\begin{enumerate}
	\item Fix $\Rey$. Compute the optimal background method bound over \mbox{2.5-D} velocity fields whose potentials $\varphi$ and $\psi$ are even in $z$.\\
	\item For $h(z)$ defined by~\cref{eq: h} using the optimal $a$ and $\zeta(z)$ found in step~(i), compute $R_o$ and $R_\varphi$ by solving the minimization problems that define them in~\cref{eq: 3 mins}. There is no need to compute $R_e$ because $R_e=Re$ by construction.\\
	\item Evaluate $\chi$ by the formula in~\cref{eq: chi}. If $\chi\le1$, the bound computed in step~(i) coincides with the optimal background method bound over \mbox{3-D} velocity fields.\\
\end{enumerate} 

Details of our own implementation of this procedure are described in \cref{sec: computational} and \cref{app: computational details}. For step~(i), our optimal background method computation is based on the fourth formulation~\cref{eq: UB opt spectral eigen}. If the criterion in step~(iii) fails to hold, meaning $\chi>1$, one cannot yet conclude that the bound from step~(i) coincides with the optimal background method bound over \mbox{3-D} fields. An option in this case is to directly check that the $a$ and $\zeta$ found in step~(i) satisfy the \mbox{3-D} spectral constraint, which would show that the bound indeed holds for \mbox{3-D} flows. Another option, which is potentially simpler but makes the bound at least slightly worse, is described in the next subsection.

\subsection{Restricting to background profiles that satisfy Busse's criteria}
\label{sec: restricted background profiles}

If an optimal background method computation over \mbox{2.5-D} fields yields $a$ and $\zeta$ for which the $\chi\le1$ criterion fails, one can repeat the bounding computation with the criterion added as a constraint. This latter \mbox{2.5-D} computation yields $a$ and $\zeta$ for which computations over \mbox{2.5-D} and \mbox{3-D} fields must coincide; thus, it gives a bound that must apply to \mbox{3-D} flows. We formulate a slightly different version of Busse's criterion to enforce as a constraint, in which $\beta$ appears linearly rather than inside of square roots.

Recall that the desired coincidence of \mbox{2.5-D} and \mbox{3-D} optima follows if $R_e$ is the smallest of the three quantities on the right-hand side of~\cref{eq: pre main result}. In other words, we need there to exist $\beta\in(0,1)$ such that $R_e\le \sqrt\beta R_o$ and $R_e\le \sqrt{1-\beta}R_\varphi$. The analysis in \cref{app: sub functional derivation} leading to~\cref{eq: pre main result} also gives, along the way, a very similar criterion in which $\beta$ appears linearly. In particular, the minimum of $\funcr[\vw]$ over \mbox{3-D} fields coincides with $R_e$ if
\begin{equation}
R_e\le \min_{\substack{\vw\in\Hdot_{2.5D}\\\varphi,\psi\text{ odd in }z}}\frac{N_o[\vw]}{\left|D_o[\vw]\right|} \quad \text{and} \quad
R_e\le \min_{\substack{\vw\in\Hdot_{2D}\\\psi=0}}\frac{N_\varphi[\vw]}{\left|D_\varphi[\vw]\right|},
	\label{eq: R_e min alternate}
\end{equation}
where the numerator and denominator functionals are defined in \cref{numerator labeling,triag inequality for lower bound} of \cref{app: sub functional derivation}. Rearranging the criterion~\cref{eq: R_e min alternate} gives an equivalent constraint,
\begin{equation}
	\label{eq: new constrained problem}
	\begin{split}
		N_o[\vw]-\Rey\,D_o[\vw]\geq 0&
		\quad \forall\,\vw\in\Hdot_{2.5D}\text{ with }\varphi,\psi\text{ odd in }z, \\
		N_\varphi[\vw]-\Rey\,D_\varphi[\vw]\geq 0&
		\quad \forall\,\vw\in\Hdot_{2D}\text{ with }\psi=0,
	\end{split}
\end{equation}
where $\beta$ appears linearly in $N_o$ and $N_\varphi$. To get~\cref{eq: new constrained problem} from~\cref{eq: R_e min alternate} we have used the fact that $R_e=\Rey$ in the present context of the background method (cf.~\cref{sec: primitive}), and we have removed the absolute values on $D_o$ and $D_\varphi$ by the same reasoning that precedes~\cref{eq: spectral abs}.

When carrying out an optimal background method computation over \mbox{2.5-D} fields, one can include the two additional constraints in~\cref{eq: new constrained problem} and optimize over $\beta\in(0,1)$ along with~$a$ and~$\zeta$. Simultaneous optimization of these parameters is not convex because $\beta$ and $a$ multiply each other, but the global optimum can still be found using a branch-and-bound algorithm, as described further in \cref{app: restricted background fields optimal upper bounds}. The resulting bounds will be guaranteed to apply to \mbox{3-D} flows. In cases where the $\chi\le1$ criterion would have been violated without these additional constraints, the constraints lead to bounds that are at least slightly worse because~$a$ and~$\zeta$ have been further constrained such that $\chi=1$. Our numerical implementation of these additional constraints is described in \cref{app: restricted background fields optimal upper bounds}, and results for Couette flow are reported in~\cref{sec: Couette}.

\section{Optimal bounds for Waleffe flow and Couette flow via dimension reduction}
\label{sec: computational}

In this section, we report bounds on dissipation computed for Waleffe flow and plane Couette flow. Both models have governing equations and boundary conditions that are symmetric under~\cref{eq: sym}, so Busse's criterion can be applied to the background method as described in \cref{sec: primitive}. Bounding computations are performed over \mbox{2.5-D} velocity fields and then shown to coincide with bounds over \mbox{3-D} fields. In the case of Waleffe flow, applicability of the bounds in \mbox{3-D} is confirmed using Busse's criterion: the $\chi\le1$ condition holds at each $\Rey$ where we computed bounds, and extrapolation of $\chi$ suggests that the condition continues to hold at larger $\Rey$. In the case of plane Couette flow, $\chi\le1$ ceases to hold once $\Rey$ exceeds a moderate value. Above this $\Rey$ value, we check directly that the spectral constraint holds also for non-zero streamwise wavenumbers. We also compute slightly worse bounds using the procedure of \cref{sec: restricted background profiles}, where the $\chi\le1$ condition is enforced as a constraint on the choice of $a$ and $\zeta$ in the background method.

For our computational implementations of the optimal background method and related variational problems, we formulate semidefinite programs (SDPs)---a standard type of convex optimization problem for which many numerical solvers are available. All past computations of optimal background method bounds have used either this SDP approach or an approach based on Euler--Lagrange equations \citep[see][]{fantuzzi_background_2021}. In the SDP approach, a quadratic variational problem reduces to an SDP after the vector field $\vw$ and background profile $\zeta$ are expanded
using finite bases. Computations can be repeated with larger bases until the resolution is sufficient for bounds to converge. Bounds reported in this section are converged to at least four significant digits, as further described in \cref{app: computational details}.

To automate the conversion of a quadratic variational problem to an SDP, we used the software QUINOPT 2.2 \citep{fantuzzi_semidefinite_2016, fantuzzi_optimization_2017}, which in turn used YALMIP R20210331 \citep{Lofberg2004} and MATLAB. To solve the resulting SDP we used Mosek 9.3 \citep{mosek}. The main advantage of the SDP approach is ease of implementation. A drawback is that one must choose a finite set of wavevectors for which to enforce the spectral constraint. This amounts to fixing periods of the domain in the streamwise and spanwise directions. One can repeat the bounding computations for different domains, but here we simply fix a domain. However, in the approach based on Euler--Lagrange equations it is possible to admit a all wavevectors, and it is easier to push computations to larger $\Rey$ values. For our present investigation, moderate $\Rey$ values suffice.

The variational problems formulated previously require further manipulation in order to be amenable to QUINOPT, including Fourier transforming in the periodic directions to obtain integrals over only the wall-normal direction. The formulation used to compute optimal bounds over \mbox{2.5-D} fields is derived in \cref{app: background method numerical scheme derivation}. The formulations used to compute the minima $R_o$ and $R_\varphi$ defined by~\cref{eq: 3 mins} are derived in \cref{app: minimizing restricted functionals}.

\subsection{Waleffe flow}
\label{sec: Waleffe}

The Waleffe flow configuration \citep{waleffe_self-sustaining_1997} is a version of Kolmogorov flow with stress-free walls and a forcing profile that is half a period of a sine function. Following \citeauthor{waleffe_self-sustaining_1997}, we choose the dimensional velocity scale $\mathsf{U}$ to be the root-mean-squared velocity of the laminar flow, but for the length scale we have chosen the layer's full thickness $d$ rather than its half-thickness. In the resulting dimensionless equations~\cref{eq: NSE}, the forcing profile is $f(z)=\Rey^{-1}\pi^2\sqrt{2}\sin(\pi z)$, and the laminar profile is $U(z)=\sqrt{2}\sin(\pi z)$. The dimensionless laminar flow has dissipation ${\left\langle {U'}^2\right\rangle}=\pi^2$.

In Waleffe flow, as in channel flow driven by a fixed pressure gradient \citep{Constantin1995a}, dissipation is maximized by its laminar value. This can be seen by choosing $a=2$ and $\zeta=2U$, so that $\funcq$ defined by~\cref{eq: Q} becomes $\funcq[\vw]=\big\langle{U'}^2 \big\rangle - \left\langle|\nabla\vw|^2 \right\rangle$. Maximizing over $\vw$ gives
\beq
\overline{\left\langle|\nabla\vu|^2\right\rangle}\le \big\langle{U'}^2\big\rangle,
\label{eq: waleffe UB}
\eeq
which must be the optimal upper bound~\cref{eq: UB opt} because it is saturated by the laminar state. Note also that the total velocity field $\vu$ obeys~\cref{eq: NSE}, and its kinetic energy evolves as
\beq
\ddt\tfrac12\left\langle|\vu|^2\right\rangle =
-\tfrac{1}{\Rey}\left\langle|\nabla\vu|^2\right\rangle + \langle f u_1\rangle.
\label{eq: ddt KE waleffe}
\eeq
The left-hand side of~\cref{eq: ddt KE waleffe} time-averages to zero, as explained preceding~\cref{eq: V identity}, so the mean dissipation is balanced by the mean work performed by the force $f(z)$,
\beq
\overline{\left\langle|\nabla\vu|^2\right\rangle} = \Rey\,\overline{\langle f u_1\rangle}.
\eeq
An interpretation of why both quantities are maximized by the laminar state is that this flow perfectly aligns the flow direction with the force direction. Here, we report our bounds on mean dissipation in terms of the ratio
\beq
\eps = \frac{1}{\Rey}\frac{\big\langle{U'}^2\big\rangle}{~\overline{\left\langle|\nabla\vu|^2\right\rangle}~}
=\frac{1}{\Rey}\frac{\big\langle fU\big\rangle}{~\overline{\left\langle fu_1\right\rangle}~},
\label{eq: ff waleffe}
\eeq
which is a kind of friction coefficient. Such a ratio is unaffected by whether or how $\vu$ has been non-dimensionalized. The laminar upper bound~\cref{eq: waleffe UB} on mean dissipation gives the lower bound $\tfrac{1}{\Rey}\le \eps$, and the lower bounds on mean dissipation that we report now give upper bounds on $\eps$.

We have computed optimal lower bounds on mean dissipation in Waleffe flow. These complement, but are not equivalent to, upper bounds from \citet{Rollin2011} on the ratio between dissipation and the $3/2$ power of kinetic energy for the same model. In our bounding computations the domain is fixed to have streamwise and spanwise periods of $\Gamma_x=\Gamma_y=2\pi$, which we have confirmed is large enough for bounds to closely approximate their large-domain limits (cf.\ the end of \cref{app: background method numerical scheme derivation}) at all but the smallest $\Rey$ values. The criterion of \cref{sec: primitive} based on Busse's theorem can only be applied in the large-domain limit because part of the theorem's proof in \cref{app: euler lagrange} assumes that all wavenumbers are admissible in the periodic directions. For our fixed domain, the energy method shows that the laminar flow is globally stable if $\Rey\le \Rey_E\approx 6.88$, and $\Rey_E$ is also the value above which our upper bounds on dissipation depart from its laminar value. This energy stability value agrees with the approximation $\Rey_E\approx7$ reported by \citet{doering2003energy} for the same fixed domain.

\begin{figure}
\centering
\hspace{-3.5pt}
\includegraphics[width=2.855in]{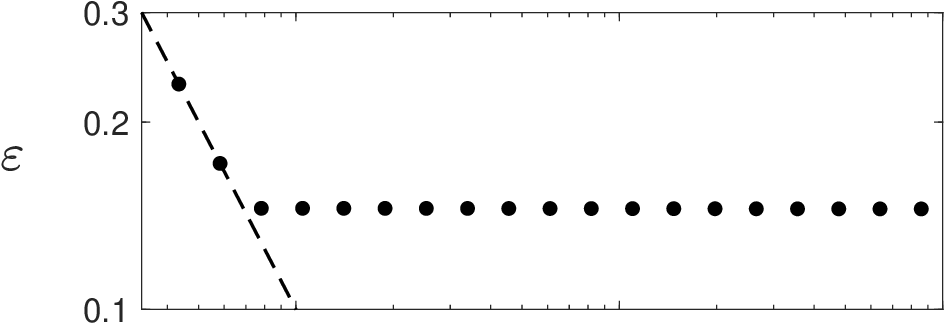}\\[4pt]
\includegraphics[width=3in]{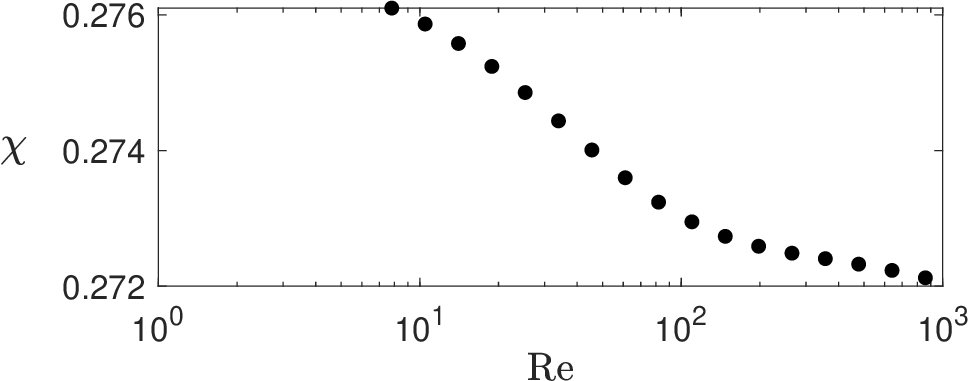}
\caption{(a) Optimal lower bounds on mean dissipation in Waleffe flow, plotted as upper bounds ($\bullet$) on the friction coefficient $\varepsilon$ defined for this model by~\cref{eq: ff waleffe}, along with the optimal lower bounds ($\dashedrule$) on $\eps$ that take the laminar value $1/\Rey$. (b) Confirmation of the $\chi\le1$ criterion, which implies that the bounds apply to all \mbox{3-D} flows despite being computed over \mbox{2.5-D} velocity fields.}
\label{fig: waleffe}
\end{figure}

\Cref{fig: waleffe}(a) shows our optimal upper bounds on the friction coefficient $\eps$ in \mbox{3-D} Waleffe flow, which asymptote to roughly 0.145. These have been computed by bounding mean dissipation below at various fixed $\Rey$ values using the procedure of \cref{sec: primitive}. To compute $R_e$ in step~(i) of this procedure, we use the formulation in \cref{app: background method numerical scheme derivation}. For the $a$ and $\zeta$ found in step~(i), values of $R_o$ and $R_\varphi$ are computed in step~(ii) using the formulations in \cref{app: minimizing restricted functionals}. \Cref{fig: waleffe}(b) shows the values of $\chi$ calculated in step~(iii) to check the $\chi\le1$ criterion~\cref{eq: chi}. These values are far smaller than unity; therefore, the bounds in \cref{fig: waleffe}(a), which were computed over \mbox{2.5-D} fields, indeed apply to \mbox{3-D} flows. Furthermore, the $\chi$ values in \cref{fig: waleffe}(b) appear to approach a constant, suggesting that the $\chi\le1$ criterion holds for all $\Rey$. If so, the optimal bound for \mbox{3-D} flows at large $\Rey$ is the asymptote of the bounds in \cref{fig: waleffe}(a).

\subsection{Plane Couette flow}
\label{sec: Couette}

Plane Couette flow has no body forcing and is driven by relative motion of the boundaries. We let the dimensionless boundary conditions be $\vu=\pm\tfrac12\vxhat$ at $z=\pm\tfrac12$, so the laminar profile is $U(z)=z$. In contrast to Waleffe flow, the laminar state of Couette flow minimizes the mean dissipation rather than maximizing it. Indeed, the lower bound
\beq
\big\langle{U'}^2\big\rangle \le
\overline{\left\langle|\nabla\vu|^2\right\rangle}
\label{eq: couette LB}
\eeq
follows from~\cref{eq: diss identity 1}, whose middle right-hand term is zero in Couette flow since it is equal to $-2\langle U''v_1\rangle$, and $U''=0$ here. In the context of the background method, this is the lower bound found with $a=0$ and $\zeta=0$. The bound is sharp because it is saturated by the laminar state. Note also that the kinetic energy of $\vu$ evolves as
\beq
\frac{\rm d}{{\rm d}t}\frac{Re}{2}\left\langle|\vu|^2\right\rangle =
-\left\langle|\nabla\vu|^2\right\rangle + 
\frac{1}{2}\frac{1}{\Gamma_x\Gamma_y}\int\left( \partial_zu_1\big|_{z=\tfrac12}+\partial_zu_1\big|_{z=-\tfrac12} \right)
{\rm d}x{\rm d}y.
\label{eq: ddt KE couette}
\eeq
The left-hand derivative vanishes in the infinite-time average, so mean dissipation is balanced by mean work by the boundaries on the flow,
\beq
\overline{\left\langle|\nabla\vu|^2\right\rangle} =  
\frac{1}{\Gamma_x\Gamma_y}\overline{\int\tfrac12
\left( \partial_zu_1\big|_{z=\tfrac12}+\partial_zu_1\big|_{z=-\tfrac12} \right) 
{\rm d}x{\rm d}y}.
\eeq
For Couette flow we follow \citet{plasting_improved_2003} and others in defining a friction coefficient as
\beq
\eps = \frac{1}{\Rey}\frac{\overline{\left\langle|\nabla\vu|^2\right\rangle}~}{\big\langle{U'}^2\big\rangle}
= \frac{1}{\Rey}\frac{\dfrac{1}{\Gamma_x\Gamma_y}\overline{\displaystyle\int
\left(\partial_zu_1\big|_{z=\tfrac12}+\partial_zu_1\big|_{z=-\tfrac12} \right) 
{\rm d}x{\rm d}y}~}{U'\big(\tfrac12\big)+U'\big(-\tfrac12\big)},
\label{eq: ff couette}
\eeq
which is a ratio unaffected by non-dimensionalization. Note that the dissipation ratio defining $\eps$ for Couette flow is inverse to the definition~\cref{eq: ff waleffe} for Waleffe flow, where the laminar value is in the numerator. The laminar lower bound~\cref{eq: couette LB} on mean dissipation in Couette flow gives the lower bound $\tfrac{1}{\Rey}\le \eps$, which is the same for Waleffe flow, and upper bounds on mean dissipation that we report later give upper bounds on $\eps$.

In our bounding computations, we fix streamwise and spanwise periods of $\Gamma_x=\Gamma_y=2\pi$. As in Waleffe flow, this approximates the large-domain limit of bounds well. In this domain, the energy method guarantees global stability of the laminar flow when $\Rey\le\Rey_E\approx 82.74$, above which our bounds depart from the laminar dissipation value. The energy stability threshold among all spanwise periods is $\Rey_E\approx 82.66$, which occurs for $\Gamma_y\approx2.016$ or a multiple thereof \citep{josephstability}, so that is the $\Rey$ above which the bounds of \citet{plasting_improved_2003} depart from the laminar value.

\begin{figure}
\centering
\hspace{-11.5pt}
\includegraphics[width=3.025in]{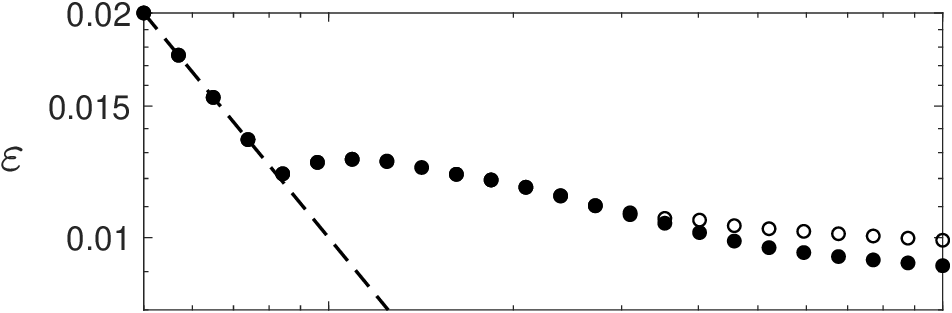}\\[6pt]
\includegraphics[width=3in]{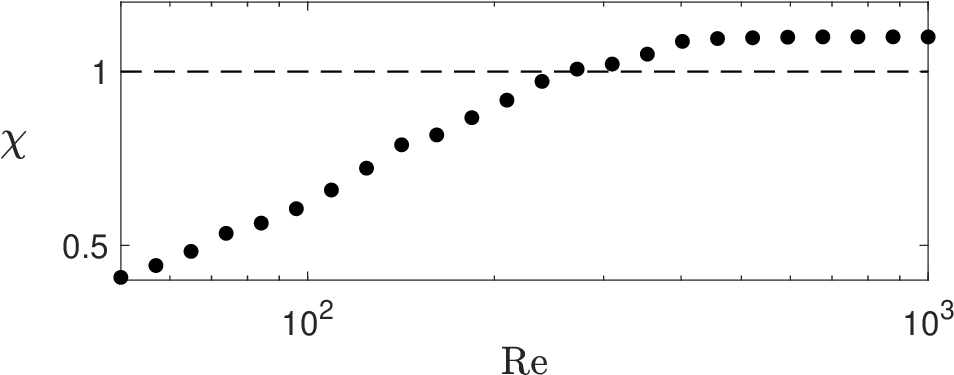}
\caption{(a) Optimal upper bounds on mean dissipation in Couette flow, computed over \mbox{2.5-D} velocity fields with no additional constraints ($\bullet$) and with constraints enforcing Busse's criterion ($\circ$). These are plotted as upper bounds on the friction coefficient $\varepsilon$ defined for this model by~\cref{eq: ff couette}, along with the optimal lower bounds ($\dashedrule$) on $\eps$ that take the laminar value $1/\Rey$. The large-$\Rey$ asymptotes are approximately 0.0086 and 0.0097, respectively. (b) Values of $\chi$ for the bounds computed without enforcing Busse's $\chi\le1$ criterion, which violate the criterion above $\Rey\approx254$.}
\label{fig: couette}
\end{figure}

\Cref{fig: couette}(a) shows upper bounds on the friction coefficient $\varepsilon$ in Couette flow computed over \mbox{2.5-D} velocity fields. The solid symbols are optimal bounds computed over \mbox{2.5-D} fields in the same way as the bounds reported previously for Waleffe flow. These bounds do not reach large enough $\Rey$ to give a precise asymptote, but they are consistent with the asymptote of $\varepsilon\lesssim0.008553$ estimated by \citet{plasting_improved_2003} based on their computations for $\Rey$ up to $7\times10^4$. Because the implementation of \citeauthor{plasting_improved_2003} was based on Euler--Lagrange equations, rather than SDPs, they were able to reach larger $\Rey$ and to enforce the spectral constraint for all spanwise wavenumbers. Nonetheless, our computations suffice to investigate the coincidence of bounds computed over \mbox{2.5-D} and \mbox{3-D} fields.

Since the bounds represented by solid symbols in \cref{fig: couette}(a) were computed over \mbox{2.5-D} fields, it remains to confirm that they apply to \mbox{3-D} flows. First we try to show this using the procedure of \cref{sec: primitive}, for which the computed bounds constitute step~(i). The last step of the procedure gives the $\chi$ values shown in \cref{fig: couette}(b). The $\chi\le1$ condition is satisfied only when $\Rey\lesssim254$, so only at these small $\Rey$ does Busse's criterion guarantee that the solid symbols in \cref{fig: couette}(a) are bounds for \mbox{3-D} Couette flow. At larger $\Rey$ values, one option is to carry out \mbox{2.5-D} bounding computations with additional constraints that enforce Busse's criterion, as proposed in \cref{sec: restricted background profiles}. We have implemented these computations as described in \cref{app: restricted background fields optimal upper bounds}. The resulting bounds, which are guaranteed to apply to \mbox{3-D} flows, appear as hollow symbols in \cref{fig: couette}(a). The asymptote of these bounds is roughly $\varepsilon\lesssim 0.009$, as estimated by fitting the $c_i$ parameters in $\varepsilon\approx c_0+c_1\Rey^{-c_2}$ to the hollow symbols. This asymptotic bound is worse than the value of 0.008553 from \citet{plasting_improved_2003}, but it is better than the previous best value of  $0.0109$ from \citet{nicodemus_background_1998}.

\begin{figure}
	\centering
	\includegraphics[width=3.75in]{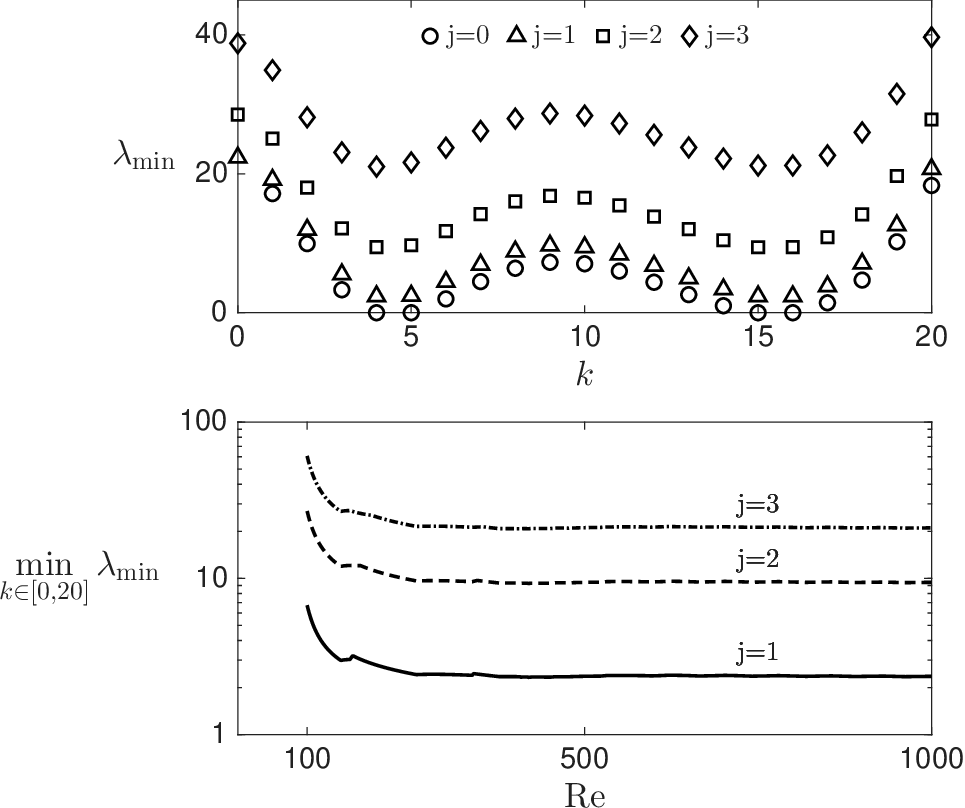}
	\caption{(a) Minimum eigenvalues $\lambda_{\min}(j,k)$ of the spectral constraint eigenproblem~\cref{eq: fourier transformed eigenvalue problem} in the $\Rey=1000$ case, computed by solving~\cref{eq: vorticity/velocity EVP}. (b) Minimum of $\lambda_{\min}$ over streamwise wavenumbers $k$ at various $\Rey$ for spanwise wavenumbers $j=1,2,3$.}
	\label{fig: spectralcouette}
\end{figure}

We have confirmed that all solid symbols in \cref{fig: couette}(a) are indeed bounds for \mbox{3-D} flows when $\Rey\gtrsim254$, despite the $\chi\le1$ criterion failing, by directly checking the spectral constraint for \mbox{3-D} fields. One formulation of the spectral constraint requires the eigenproblem~\cref{eq: fourier transformed eigenvalue problem} to have only non-negative eigenvalues for all admissible wavevectors $\wn=(j,k)$. Since the spectrum of eigenvalues is real and bounded below, we need only find the minimum eigenvalue $\lambda_{\min}$ at each wavevector. With the present periods of $\Gamma_x=\Gamma_y=2\pi$, the admissible $(j,k)$ include all pairs of non-negative integers except $(0,0)$. The $\lambda_{\min}\ge0$ constraints for $j=0$ are already enforced by the \mbox{2.5-D} bounding computations, but  the constraints for \mbox{$j\ge1$} remain to be checked \emph{a posteriori}.  At each $\Rey$, with $a$ and $\zeta$ obtained from the \mbox{2.5-D} bounding computation, we computed $\lambda_{\min}$ for various $(j,k)$ using the software Dedalus 3.0.2 \citep{Burns}. These computations implemented an eigenproblem that is equivalent to~\cref{eq: fourier transformed eigenvalue problem}, but written using the wall-normal velocity $w_3$ and wall-normal vorticity $\eta=\partial_xw_1-\partial_yw_1$, whose Fourier transforms $\hat w_3(z)$ and $\hat\eta(z)$ must satisfy
\begin{equation}
\label{eq: vorticity/velocity EVP}
\begin{split}
2\lambda\hat\eta+2\left(a-1\right)\left(\tfrac{\mathrm{d}^2}{\mathrm{d}z^2}-|\wn|^2\right)\hat\eta+i\Rey\left(aU+\zeta\right)^{\prime}k \hat{w}_3 &= 0, \\[4pt]
2\lambda\left(\tfrac{\mathrm{d}^2}{\mathrm{d}z^2}-|\wn|^2\right) \hat{w}_3
+2\left(a-1\right)\left(\tfrac{\mathrm{d}^2}{\mathrm{d}z^2}-|\wn|^2\right)^2 \hat{w}_3 \hspace{40pt} \\[-2pt]
+\,i\Rey\Bigl[\bigl(2j\tfrac{\mathrm{d}}{\mathrm{d}z}\hat{w}_3+k\eta\bigr)\left(aU+\zeta\right)^{\prime}
+j\hat{w}_3\left(aU+\zeta\right)^{\pprime}\Bigr] &= 0.
\end{split}
\end{equation}
The no-slip boundary conditions require that $\hat w_3=\hat w_3'=\hat\eta=0$ at $z=\pm1/2$. For each $\Rey$, the $\zeta(z)$ and $a$ in \cref{eq: vorticity/velocity EVP} come from the bounding computation over \mbox{2.5-D} fields using QUINOPT. A Legendre basis was used to discretize $z$ in Dedalus to match the basis in which QUINOPT returns $\zeta(z)$. At each $\Rey$ where we computed bounds, we then solved~\cref{eq: vorticity/velocity EVP} to find $\lambda_{\min}(j,k)$ at all admissible pairs of $j\in[0,3]$ and $k\in[0,20]$. \Cref{fig: spectralcouette}(a) shows the resulting $\lambda_{\min}$ values in the $\Rey=1000$ case. Here $\lambda_{\min}=0$ when $(j,k)=(0,4)$, $(0,5)$, $(0,15)$ or $(0,16)$, meaning these are the wavevectors at which the spectral constraint is active, which is consistent with figure 3 of \citet{plasting_improved_2003}. All other $\lambda_{\min}(j,k)$ are strictly positive. Our computations at other $\Rey$ give similar $\lambda_{\min}$ trends, including the fact that $\lambda_{\min}$ increases monotonically with $j$ at every $k$.

For evidence that optimal bounds over \mbox{2.5-D} and \mbox{3-D} fields continue to coincide as $\Rey\to\infty$, we aim to extrapolate the finding that $\lambda_{\min}(j,k)>0$ when $j\ge1$. To do this, we minimize $\lambda_{\min}(j,k)$ over $k$ and examine the dependence on $\Rey$. \Cref{fig: spectralcouette}(b) shows the results for $j=1,2,3$. Each curve asymptotes to a strictly positive value as $\Rey$ is raised, and the asymptotic value increases quickly with $j$. If this trend continues, it justifies the claim of \citet{plasting_improved_2003} that their \mbox{2.5-D} bounds give the optimal bounds for \mbox{3-D} Couette flow.

In the case of Couette flow, we can compare the computational cost of checking the \mbox{3-D} spectral constraint with the cost of checking the $\chi\le1$ criterion. Both are less expensive than the first step of computing optimal bounds over \mbox{2.5-D} fields. Checking the \mbox{3-D} spectral constraint requires computing $\lambda_{\min}$ for various $(j,k)$ pairs, whereas checking the $\chi\le1$ criterion requires computing $R_o$ and $R_\varphi$. With $\Rey=1000$, for instance, the time to compute $\lambda_{\min}$ for a single $(j,k)$ pair using Dedalus was roughly half of the time needed to compute $R_o$ and $R_\varphi$ using QUINOPT, so checking the spectral constraint for any reasonable number of $(j,k)$ pairs would be significantly more expensive than checking our $\chi\le1$ criterion.

\section{Discussion and conclusions}
\label{sec:conclusion}

We have studied the variational problems that arise when using the background method to bound mean dissipation above or below in planar shear flows. Our main theoretical contribution is a criterion for confirming that the variational problems have the same optimum over \mbox{2.5-D} (streamwise-invariant) fields and \mbox{3-D} fields. This criterion relies on a theorem of \citet{busse_property_1972} for the energy stability problem, which can be seen as a special case of the background method. Busse's theorem and our criterion apply only to planar shear flow models that are symmetric under a rotation about a spanwise axis that swaps the two walls. We have also derived four equivalent formulations of the optimal background method, none of which is new, and in \cref{app: background sym} we have used a standard argument to show that one-dimensional background profiles suffice for optimal bounds.

We have reported bounds on dissipation that are optimal within the background method for both Waleffe flow and plane Couette flow up to moderate $\Rey$ values. For Waleffe flow we computed lower bounds, and for Couette flow we computed upper bounds, both of which correspond to upper bounds on a friction coefficient when such a coefficient is defined reasonably for each model. These friction coefficients are bounded below by their laminar values of $1/\Rey$, whereas our upper bounds approach constants as $\Rey$ is raised. The bounds for Couette flow are consistent with the more extensive computations of \citet{plasting_improved_2003}. All of our bounds were computed over \mbox{2.5-D} velocity fields, after which we showed that they must coincide with results over \mbox{3-D} fields. For Waleffe flow our new criterion confirmed this coincidence at all $\Rey$ where bounds were computed, and extrapolation suggests the criterion holds at all other $\Rey$ also. For Couette flow our criterion holds only when $\Rey\lesssim254$, so up to $\Rey=1000$ we instead verified directly that the \mbox{2.5-D} optima satisfy additional spectral constraints implying that they are also \mbox{3-D} optima. Extrapolation suggests that the spectral constraints continue to hold at larger $\Rey$, thus supporting the assumption of \citet{plasting_improved_2003} that their optimal bounds computed over \mbox{2.5-D} fields indeed apply to \mbox{3-D} flows. The computational cost of checking our new criterion was significantly lower than the cost of checking the spectral constraint in every case where we did both.

For shear flow models that lack the symmetry needed to apply our criterion, or that are not planar, it is an open challenge to find criteria that can verify coincidence of optima over \mbox{2.5-D} and \mbox{3-D} fields. This is true for the energy stability problem as well as the background method. Busse's argument, a version of which is given in \cref{app: proof of busse theorem}, relies fundamentally on the rotational symmetry. In particular, in \cref{app: sub functional derivation}, the numerator and denominator of the $\funcr$ functional are each decomposed into terms that depend on three different projections of the velocity field. This decomposition will not occur without the rotational symmetry because there will be additional terms. In such cases, Busse's approach cannot show coincidence of \mbox{2.5-D} and \mbox{3-D} optima, but it can give upper bounds on how far apart the two optima can be. This has been done for the energy stability problem of channel flow \citep{kaiser_bounds_2001}, and similar arguments for the background method might be able to estimate how far apart the optimal bounds can be when computed over \mbox{2.5-D} and \mbox{3-D} fields.

Despite the lack of theoretical guarantees for other shear flows, computations for some of these other models yield critical eigenmodes of the energy stability problem that are streamwise-invariant rather than fully \mbox{3-D}. This is the case for channel flow and Taylor--Couette flow at most radius ratios \citep{joseph_stability_1969} and various other models \citep[e.g., reported by][]{xiong2019conjecture}. However, critical energy eigenfunctions are fully \mbox{3-D} for pipe flow \citep{joseph_stability_1969} and for Taylor--Couette flow when the inner cylinder is much smaller than the outer one \citep{kumar_geometrical_2022}. Optimal background method bounds have been computed for some of these models, always assuming---but usually not verifying---that it suffices to compute over \mbox{2.5-D} fields \citep{nicodemus_background_1998,nicodemus_background_1998-1,kerswell_new_2001,plasting_improved_2003,fantuzzi_bounds_2018,arslan_bounds_2021,kumar_analytical_2022,kumar_geometrical_2022}. Verification that such bounds indeed apply to all \mbox{3-D} flows calls for a better theoretical understanding of when optimal velocity fields must have certain symmetries, both in energy stability analysis and in the background method more generally.

\section*{Acknowledgements}
We thank Elizabeth Carlson, Giovanni Fantuzzi, Rich Kerswell and Anuj Kumar for some very helpful discussions. Both authors were supported by Canadian NSERC Discovery Grants Program awards RGPIN-2018-04263, RGPAS-2018-522657 and RGPIN-2025-06823.

\appendix
\crefalias{section}{appsec}
\crefalias{subsection}{appsec}
\crefalias{subsubsection}{appsec}

\section{Symmetries of the optimal background field}
\label{app: background sym}

Symmetries of the governing model allow symmetries to be imposed on the background field without worsening bounds. \Cref{app: zeta 1D} proves that it suffices to consider one-dimensional background fields in planar shear flows, and \cref{app: zeta odd} proves that odd background profiles suffice when the model has an additional symmetry. The arguments apply equally to upper and lower bounds.

\subsection{Optimality of one-dimensional background fields}
\label{app: zeta 1D}

Consider a planar shear flow as described at the start of \cref{sec: problem setup}, where deviation from the laminar flow is governed by~\cref{eq: NSE v}. Suppose one considers a \mbox{3-D} background field $\boldsymbol{\zeta}(\vx)$, in which case the auxiliary function~\cref{eq: V} is replaced by
\begin{equation}
	V[\vw]=\Rey\left\langle\tfrac{a}{2}|\vw|^2- \boldsymbol{\zeta}\cdot\vw\right\rangle. 
\end{equation}
Repeating the calculations after~\cref{eq: V} with this more general $V$, we find that $\boldsymbol{\zeta}(\vx)$ must be divergence-free to avoid a pressure term in the functional $\funcq$, in which case $\funcq$ is
\begin{equation}
	\begin{split}
		\funcq[\vw;\boldsymbol{\zeta}(\vx),a]
		&= \big\langle U'^2+2U'\partial_zw_1-(a-1)|\nabla\vw|^2-a\Rey U'w_1w_3\big\rangle \\
		&\quad -\Rey\left\langle \boldsymbol{\zeta}\cdot \left(- \vw\cdot\nabla\vw - U\partial_x\vw - U'w_3\hat{\vx} + \tfrac{1}{\Rey}\lap\vw \right)\right\rangle.
	\end{split}
	\label{eq: Q general zeta}
\end{equation}
For the case of upper bounds, a chosen pair of $\boldsymbol{\zeta}$ and $a$ can be used to prove a bound $B$ if and only if
\begin{equation}
	\funcq[\vw;\boldsymbol{\zeta}(\vx),a]\leq B \quad \forall\vw\in\mathcal{H}_{3D}.
	\label{eq: B general zeta}
\end{equation}
The governing model is invariant under every translation in the periodic $x$ and $y$ directions, so if $\vw(\vx)$ is in $\mathcal{H}_{3D}$, then so are all translations of $\vw(\vx)$. Therefore,~\cref{eq: B general zeta} is equivalent to the same condition holding for all translations of $\zeta(\vx)$,
\begin{equation}
	\funcq[\vw;\boldsymbol{\zeta}\left(x+\gamma_x,y+\gamma_y,z\right),a]
	\leq B\quad \forall\vw\in\mathcal{H}_{3D}
	\label{eq: B general zeta translated}
\end{equation}
for all $(\gamma_x,\gamma_y)$. Averaging both sides of~\cref{eq: B general zeta translated} over all translations gives
\begin{equation}
	\label{eq: averaging}
	\frac{1}{L_xL_y}\int_0^{L_x} \int_0^{L_y}\funcq[\vw;\boldsymbol{\zeta}\left(x+\gamma_x,y+\gamma_y,z\right),a]\,\mathrm{d}\gamma_x\mathrm{d}\gamma_y \le B.
\end{equation}
The integral in~\cref{eq: averaging} acts only on $\boldsymbol{\zeta}$, which appears linearly in $\funcq$. Therefore~\cref{eq: averaging} is equivalent to
\begin{equation}
	\funcq[\vw;\widetilde{\boldsymbol{\zeta}}(z),a]\leq B \quad \forall\vw\in\mathcal{H}_{3D},
	\label{eq: B general zeta averaged}
\end{equation}
where $\widetilde{\boldsymbol{\zeta}}(z)$ denotes the average of $\boldsymbol{\zeta}(\vx)$ over its translations in $x$ and $y$. The components of the averaged background field are $(\widetilde\zeta_1(z),\widetilde\zeta_2(z),0)$, where the third component vanishes because $\boldsymbol{\zeta}$ is divergence-free and satisfies the same impenetrability condition as~$\vw$ at the boundaries. Therefore, any bound $B$ that can be shown using $\boldsymbol{\zeta}(\vx)$ via~\cref{eq: B general zeta} can also be shown using $(\widetilde\zeta_1(z),\widetilde\zeta_2(z),0)$ via~\cref{eq: B general zeta averaged}. 

It remains to show that~\cref{eq: B general zeta averaged} holds also when $\widetilde\zeta_2=0$. For the averaged background field, the expression~\cref{eq: Q general zeta} for $\funcq$ becomes
\begin{equation}
	\begin{split}
		\funcq[\vw,\widetilde{\boldsymbol\zeta}(z),a] 
		&=\left\langle U'^2+(2U'+{\widetilde\zeta}'_1)\partial_zw_1-(a-1)|\nabla\vw|^2-\Rey\big(aU'+\widetilde\zeta_1'\big)w_1w_3\right\rangle \\
		&\quad + \left\langle \widetilde\zeta_2'\partial_zw_2-\Rey\,\widetilde\zeta_2'w_2w_3 \right\rangle.
	\end{split}
	\label{eq: Q zeta2}
\end{equation}
The second line in~\cref{eq: Q zeta2} cannot be helpful in satisfying~\cref{eq: B general zeta averaged}. For any $\vw$ in $\mathcal{H}_{3D}$, its spanwise reflection $[w_1,w_2,w_3](x,-y,z)$ is also $\mathcal{H}_{3D}$, and the two fields give the same value for the first integral in~\cref{eq: Q zeta2}, but opposite signs for the second integral. Therefore, if~\cref{eq: B general zeta averaged} holds for a background field $(\widetilde\zeta_1(z),\widetilde\zeta_2(z),0)$, then it also holds for the one-dimensional field $\widetilde\zeta_1(z)\vxhat$. The preceding arguments apply also to the case of lower bounds simply by reversing the inequalities in \cref{eq: B general zeta,eq: B general zeta translated,eq: averaging,eq: B general zeta averaged}. Thus, any upper or lower bound that can be proved with $\boldsymbol\zeta(\vx)$ can be proved with $\widetilde\zeta_1(z)\vxhat$. This justifies the restriction to background fields pointing in only the streamwise direction and varying in only the wall-normal direction.

\subsection{Optimality of odd background profiles for models with an additional symmetry}
\label{app: zeta odd}

Consider a planar shear flow as described at the start of \cref{sec: problem setup} whose governing model is symmetric under the rotation~\cref{eq: sym} that swaps the walls. As shown in \cref{app: zeta 1D}, it suffices to consider a one-dimensional background field $\zeta(z)\vxhat$, so that the expression for $\funcq$ is~\cref{eq: Q}. In the case of upper bounds, a chosen pair of $\zeta(z)$ and $a$ can be used to prove a bound $B$ if and only if
\begin{equation}
	\funcq[\vw;\zeta(z),a]\leq B \quad \forall\vw\in\mathcal{H}_{3D}.
	\label{eq: B 1D zeta}
\end{equation}
We claim that~\cref{eq: B 1D zeta} holds also for background profile $-\zeta(-z)$. To see this, note first that the symmetry~\cref{eq: sym} implies that $\mathcal{H}_{3D}$ contains $[w_1,w_2,w_3](x,y,z)$ if and only if it contains the transformed field $[-w_1,w_2,-w_3](-x,y,-z)$. Therefore,~\cref{eq: B 1D zeta} is equivalent to the same condition holding with $\vw$ replaced by $[-w_1,w_2,-w_3](-x,y,-z)$,
\beq
\funcq = \big\langle U'^2+[2U'(z)+\zeta'(z)]\partial_zw_1-(a-1)|\nabla\vw|^2-\Rey[aU'(z)+\zeta'(z)]w_1w_3\big\rangle.
\label{eq: Q trans}
\eeq
The only change from~\cref{eq: Q} to \cref{eq: Q trans} is that the coordinates of $\vw$ are $(-x,y,-z)$. Redefining $(-x,y,-z)\mapsto(x,y,z)$ and using the evenness of $U'(z)$ gives
\beq
\funcq = \big\langle U'^2+[2U'(z)+\zeta'(-z)]\partial_zw_1-(a-1)|\nabla\vw|^2-\Rey[aU'(z)+\zeta'(z)]w_1w_3\big\rangle,
\label{eq: Q trans 2}
\eeq
where now the coordinates of $\vw$ are $(x,y,z)$ as usual. We have shown that the condition~\cref{eq: B 1D zeta} is equivalent to the same condition holding for the $\funcq$ in~\cref{eq: Q trans 2}. Observe that the expression~\cref{eq: Q trans 2} is identical to the result of replacing $\zeta(z)$ by $-\zeta(-z)$ in the expression~\cref{eq: Q} for $\funcq[\vw,\zeta(z),a]$, and so \cref{eq: B 1D zeta} holds if and only if
\begin{equation}
	\funcq[\vw;-\zeta(-z),a]\leq B \quad \forall\vw\in\mathcal{H}_{3D}. 
	\label{eq: B 1D zeta trans}
\end{equation}
Averaging the inequalities in~\cref{eq: B 1D zeta} and~\cref{eq: B 1D zeta trans} gives
\begin{equation}
	\tfrac{1}{2}\big(\funcq[\vw;\zeta(z),a]+\funcq[\vw;-\zeta(-z),a]\big)\leq B  
	\quad \forall\vw\in\mathcal{H}_{3D}.
	\label{eq: B 1D zeta averaged}
\end{equation}
Because $\funcq[\vw;\zeta(z),a]$ is linear in $\zeta$, we can rewrite~\cref{eq: B 1D zeta averaged} as
\begin{equation}
	\funcq[\vw;\widetilde\zeta(z),a]\leq B  
	\quad \forall\vw\in\mathcal{H}_{3D},
	\label{eq: B 1D zeta odd}
\end{equation}
where $\widetilde\zeta(z)=\tfrac12[\zeta(z)-\zeta(-z)]$ is the odd part of $\zeta(z)$.
Thus, any bound $B$ that can be shown via~\cref{eq: B 1D zeta} with some $\zeta(z)$ can also be shown with the odd part of $\zeta(z)$ instead. Analogous arguments hold in the case of lower bounds. This justifies the restriction to odd background profiles for planar models with the symmetry~\cref{eq: sym}.

\section{Computational formulations}
\label{app: computational details}

This appendix describes how we manipulate several of the variational problems derived previously so that they can be numerically solved using the software QUINOPT. Examples of resolution and convergence are reported for the main computations of optimal bounds. For concreteness, we describe the case of upper bounds. Lower bounds require only minor modifications.

\subsection{Optimal bounds on dissipation}
\label{app: background method numerical scheme derivation}

In the case of upper bounds, our first formulation~\cref{eq: UB opt} of the optimal background method, restricted to \mbox{2.5-D} fields, may be restated as the constrained minimization
\begin{equation}
	\overline{\left\langle|\nabla\vu|^2\right\rangle}\leq\min_{\substack{a,\,\zeta}}B
	\quad\text{s.t.}\quad B-\funcq[\vw]\ge0~~\forall\,\vw\in\mathcal{H}_{2.5D},
	\label{eq: UB opt cons}
\end{equation}
where $\funcq$ is as defined in~\cref{eq: Q}. The quadratic integral constraint in~\cref{eq: UB opt cons} can be enforced separately for each spanwise wavenumber $k$ by Fourier transforming the components of $\vw(y,z)$. For the $k=0$ wavenumber this gives the spanwise average of the constraint in~\cref{eq: UB opt cons},
\beq
\int_{-\frac{1}{2}}^{\frac{1}{2}}\left[B-U'^2+(a-1)\hat{w}_1'^2+\left(2U^{\prime\prime}+\zeta''\right)\hat{w}_1\right]\mathrm{d}z \geq 0 \quad \forall\,\hat{w}_1\in\mathcal{H}_{w_1},
\label{eq: k=0 constraint}
\eeq
where $\hat{w}_1(z)$ is real, and primes denote $z$ derivatives as usual. The function space $\mathcal{H}_{w_1}$ encodes boundary conditions on $\hat{w}_1$, which are the same as on $w_1$, as well as the symmetry of $\hat{w}_1$ described later. For each $k>0$, the constraint in~\cref{eq: UB opt cons} implies
\begin{multline}
	\int_{-\frac{1}{2}}^{\frac{1}{2}}
	\Big[(a-1)\big(|\hat{w}'_1|^2+k^2|\hat{w}_1|^2+\tfrac{1}{k^2}|\hat{w}''_3|^2+2|\hat{w}'_3|^2+k^2|\hat{w}_3|^2\big)+\Rey\left(aU^\prime+\zeta'\right)\hat{w}_1\hat{w}_3\Big]\mathrm{d}z \geq 0 \\ \quad \forall~\hat{w}_1\in\mathcal{H}_{w_1}\text{ and }\hat{w}_3\in\mathcal{H}_{w_3}, 
	\label{eq: k>0 constraints}
\end{multline}
where $\mathcal{H}_{w_3}$ encodes the symmetry of $\hat{w}_3$ (cf.\ later) and its boundary conditions, which are $\hat{w}_3,\hat{w}'_3=0$ or $\hat{w}_3,\hat{w}''_3=0$ if conditions on $\vw$ are no-slip or stress-free, respectively. To derive~\cref{eq: k>0 constraints} we have eliminated the $\hat{w}_2$ terms using the relation $\hat{w}_2=(i/k)\hat{w}'_3$ that follows from Fourier transforming the divergence-free condition for \mbox{2.5-D} fields. Although the Fourier components $\hat{w}_1$ and $\hat{w}_3$ are complex in general, it suffices to enforce \cref{eq: k>0 constraints} for real functions because the constraints from the real and imaginary parts decouple and are redundant.

With its constraint decomposed for each spanwise wavenumber, the optimal bound~\cref{eq: UB opt cons} over \mbox{2.5-D} fields can be expressed as
\begin{equation}
\overline{\left\langle|\nabla\vu|^2\right\rangle}\leq\min_{\substack{a,\,\zeta}}B
\quad\text{s.t.}
\begin{array}[t]{l} \cref{eq: k=0 constraint}, \\
\cref{eq: k>0 constraints}~\forall k\in K
\end{array}
\label{eq: UB opt cons 2}
\end{equation}
for some set $K$ of wavenumbers. Although the constraint~\cref{eq: k>0 constraints} must hold for all admissible~$k$, which are the positive integer multiples of $2\pi/\Gamma_y$, only a finite number of these constraints will affect the optimum of~\cref{eq: UB opt cons}. It can be shown \emph{a priori} that the constraint is automatically satisfied for sufficiently large $k$ \citep{fantuzzi_bounds_2018}. Among smaller $k$, one can sometimes guess other ranges of $k$ values for which the constraint \cref{eq: k>0 constraints} need not be enforced when computing the right-hand optimum in~\cref{eq: UB opt cons 2}, and such guesses can be confirmed \emph{a posteriori} by checking these unenforced constraints one by one. In any case, once a finite set $K$ is chosen for which to enforce the constraint, the right-hand minimum can be computed using QUINOPT \citep{QUINOPT}.

Step~(i) of the procedure in \cref{sec: primitive} requires computing the optimal background method bound over \mbox{2.5-D} fields with potentials $\varphi$ and $\psi$ that are even in $z$. Since we have used primitive variables to formulate the constraints \cref{eq: k=0 constraint,eq: k>0 constraints}, we must determine how even potentials correspond to symmetries of $w_1$ and $w_3$, and the latter can be encoded in the function spaces $\mathcal{H}_{w_1}$ and $\mathcal{H}_{w_3}$. The field $\vw(y,z)$ is \mbox{2.5-D} but not mean-free, so expanding into its spanwise mean $\mathbf{F}(z)$ and poloidal--toroidal parts \cref{eq: polo tor} gives
\begin{align}
	w_1(y,z)&=F_1(z)+ \partial_y\psi(y,z), 
	&
	w_3(y,z)&=-\partial_y^2\varphi(y,z). 
	\label{eq: w syms}
\end{align}
Recall from \cref{sec: decoupling of mean flow} that the optimizing velocity field has mean $F_1^*(z)=\tfrac{1}{2c}\zeta(z)$, which is odd, so it suffices to enforce the constraints of~\cref{eq: UB opt cons 2} only for odd $F_1(z)$. Then~\cref{eq: w syms}, implies that $\varphi$ and $\psi$ being even (resp., odd) in $z$ corresponds to $w_1(y,z)-F_1(z)$ and $w_3(y,z)$ being even (resp., odd) in $z$. This is enforced by including only even Legendre polynomials in the expansions of $\hat{w}_1$ and $\hat{w}_3$ in the $k>0$ constraint \cref{eq: k>0 constraints}.

Implementing the minimization~\cref{eq: UB opt cons 2} in QUINOPT requires choosing finite polynomial bases for $\zeta$, $\hat{w}_1$ and $\hat{w}_3$. Shrinking the space for $\zeta$ in this way cannot decrease the right-hand minimum in~\cref{eq: UB opt cons 2}, but shrinking the spaces for $\hat{w}_1$ and $\hat{w}_3$ can lead to a smaller minimum that is not guaranteed to be an upper bound on dissipation. For the minimum to be a guaranteed upper bound---aside from numerical error---the polynomial bases of $\hat{w}_1$ and $\hat{w}_3$ must be enlarged until the minimum converges. Our computations use the Legendre polynomial basis for all three functions, with only odd terms in the basis for $\zeta$ and only even terms in the basis for $\hat{w}_1-F_1$ and $\hat{w}_3$. For $\zeta$ of various maximum polynomials degrees $N_\zeta$, convergence tests suggest that letting the maximum degree of $\hat{w}_1$ and $\hat{w}_3$ be $N_w=2N_\zeta$ approximates the large-$N_w$ limit to within five significant digits. We thus fix $N_w=2N_\zeta$ and increase $N_\zeta$ until the right-hand minimum in~\cref{eq: UB opt cons 2} converges to the optimal bound. \Cref{tbl:bounds convergence} gives some examples of how the right-hand minimum of~\cref{eq: UB opt cons 2} converges as $N_\zeta$ is raised in the case of Couette flow, and convergence was similar in the case of lower bounds for Waleffe flow. The tabulated computations become more expensive from top to bottom; the runtime on a laptop with 2.6 GHz Intel Core i7 running MATLAB2024b for QUINOPT, including the SDP solution by Mosek, ranged from several seconds to approximately one minute. For all of the bounds shown in \cref{fig: waleffe,fig: couette}, we have fixed $N_w=2N_\zeta$ and chosen $N_\zeta$ such that convergence is similar to the examples in \cref{tbl:bounds convergence}. 
\begin{table}
\begin{center}
\def~{\hphantom{0}}
\begin{tabular}{cccc}
$\Rey$  & Maximum $k$ & $N_\zeta$ & Minimum of \cref{eq: UB opt cons 2} \\[4pt]
			100 & 10 & 5 & 1.268234 \\
			&& 15 & 1.267977 \\
			&& 25 & 1.267977 \\[4pt]
			500 & 10 & 15 & 4.877071 \\
			&& 25 & 4.875818 \\
			&& 35 & 4.875817 \\[4pt]
			1000 & 20 & 25 & 9.168796 \\
			&&35 & 9.168630 \\
			&&45 & 9.168627 \\[4pt]
			2000 & 40 & 35 & 17.698797 \\
			&& 45 & 17.701393 \\
			&& 55 & 17.701393
\end{tabular}
\caption{Right-hand minima of~\cref{eq: UB opt cons 2} computed using QUINOPT with polynomial $\zeta$ of degree $N_\zeta$, and polynomials $\hat{w}_1$ and $\hat{w}_3$ of degree $N_w=2N_\zeta$. The constraint~\cref{eq: k>0 constraints} is imposed for wavenumbers $k=1,2,\ldots$ up the tabulated maximum $k$. The reported minima have been rounded to the precision shown.}
\label{tbl:bounds convergence}
\end{center}
\end{table}

Our application of Busse's theorem to the background method requires bounds to be computed in the large-domain limit because the theorem's proof assumes that all streamwise and spanwise wavenumbers are admissible (cf.\ \cref{app: euler lagrange}). To confirm that bounds computed over \mbox{2.5-D} fields with $\Gamma_y=2\pi$ are close approximations of the $\Gamma_y\gg1$ limit, we repeated the computations in \cref{tbl:bounds convergence} with a set $K$ of spanwise wavenumbers corresponding to $\Gamma_y=3\pi$. At each $\Rey$, bounds computed with the middle $N_\zeta$ resolution for the two different $\Gamma_y$ differ by 0.6\% or less.

\subsection{Computation of $R_o$ and $R_\varphi$}
\label{app: minimizing restricted functionals}

After $a$ and $\zeta$ are found in the optimal bounding computations of \cref{app: background method numerical scheme derivation}, or the analogous computations for lower bounds, one can check the criterion of \cref{sec: primitive} by computing the minima $R_o$ and $R_\varphi$ defined in~\cref{eq: 3 mins}. Whereas the formulation in~\cref{app: background method numerical scheme derivation} uses primitive variables, we use poloidal--toroidal variables to compute $R_o$ and $R_\varphi$. These values are minima of the $\funcr[\vw]$ functional over different subspaces, and simplifying the poloidal--toroidal expression~\cref{eq: R pol tor} for $\funcr$ over each subspace gives
\begin{align}
\label{eq: Ro Rphi}
R_o&=\min_{\substack{\vw\in\Hdot_{2.5D}\\\varphi,\psi\text{ odd in }z}}
\frac{\left\langle\left|\partial_{y} \lap \varphi\right|^{2}+\left|\partial_{yz}\psi\right|^{2}+\left|\partial_y^2\psi\right|^{2}\right\rangle}{\left|\left\langle h\partial_y^{2} \varphi\partial_{y} \psi\right\rangle\right|}, &
R_\varphi&=\min_{\substack{\vw\in\Hdot_{2D}\\\psi=0}}
\frac{\left\langle\left|\partial_x\lap\varphi\right|^2\right\rangle}{\left|\left\langle h\lap_2\varphi\partial_{xz}\varphi\right\rangle\right|}.
\end{align}
These minimizations of ratios can be reformulated as constrained maximizations,
\begin{subequations}
\label{eq: Ro Rphi max}
\begin{align}
	\label{eq: Ro max}
	R_o &= \max R \quad \text{s.t.} \quad \hspace{-2pt}
	\begin{array}[t]{r}
		\left\langle\left|\partial_{y} \lap \varphi\right|^{2}+\left|\partial_{yz}\psi\right|^{2}+\left|\partial_y^2\psi\right|^{2}\right\rangle - R \left\langle h\partial_y^{2} \varphi\left(\partial_{y} \psi\right)\right\rangle \ge 0 \hspace{20pt}~ \\[2pt]
		\forall \vw\in\Hdot_{2.5D}\text{ with }\varphi,\psi~\text{odd in }z,
	\end{array} \\
	\label{eq: Rphi max}
	R_\varphi &= \max R \quad \text{s.t.} \quad 
	\left\langle\left|\partial_x\lap\varphi\right|^2\right\rangle
	-R\left\langle h\lap_2\varphi\partial_{xz}\varphi\right\rangle\geq 0 \quad
	\forall \vw\in\Hdot_{2D}\text{ with }\psi = 0.
\end{align}
\end{subequations}
The argument showing that~\cref{eq: Ro Rphi} and~\cref{eq: Ro Rphi max} are equivalent is essentially the same as the argument leading from \cref{eq: spectral h} to \cref{eq: spectral Re}.

The constraint in~\cref{eq: Ro max} can be enforced separately for each spanwise wavenumber $k$. Letting $\hat{P}(z)$ and $\hat{T}(z)$ denote the Fourier transforms of $P=\partial_y\varphi(y,z)$ and $T=\partial_y\psi(y,z)$, respectively, the constraint for each spanwise wavenumber $k>0$ becomes
\begin{multline}
\label{fourier transformed optimization odd}
\int_{-\frac{1}{2}}^{\frac{1}{2}}\left[\left(a-1\right)\left(\tfrac{1}{k^2}\left|\hat{P}''\right|^2+2\left|\hat{P}'\right|^2+k^2\left|\hat{P}\right|^2+\tfrac{1}{k^2}\left|\hat{T}'\right|^2+\left|\hat{T}\right|^2\right)-\tfrac{i}{k}R\left(aU^\prime+\zeta^\prime\right)\hat{P}\hat{T}^\dagger\right]\mathrm{d}z\geq 0 \\ 
\forall \hat{P}\in\mathcal{P}_o,~\hat{T}\in\mathcal{T}_o,
\end{multline}
where we have multiplied all terms by the denominator $a-1$ of $h(z)$, which is positive in the upper bound case, so that the optimization parameters $a$ and $\zeta$ appear linearly in the constraint. Note that there is no $k=0$ constraint since $P$ and $T$ are mean-free. The spaces $\mathcal{P}_o$ and $\mathcal{T}_o$ encode oddness in $z$ and the boundary conditions, which are $\hat{P},\hat{P}^\prime,\hat{T}=0$ or $\hat{P},\hat{P}^{\pprime},\hat{T}^\prime=0$ if the conditions on $\vw$ are no-slip or stress-free, respectively. The fields $\hat{P}$ and $\hat{T}$ are complex in general, and $\hat{T}^\dagger$ denotes a complex conjugate.

The constraint in~\cref{eq: Rphi max} can be enforced for each streamwise wavenumber $j$. In this case, we let $\hat{P}(z)$ denote the Fourier transform of $P=\partial_x\varphi(x,z)$ in the $x$ direction, and the constraint for each $j>0$ becomes
\begin{equation}
\label{fourier transformed optimization}
\int_{-\frac{1}{2}}^{\frac{1}{2}}
\left[\left(a-1\right)\left(\tfrac{1}{j^2}\left|\hat{P}''\right|^2+2\left|\hat{P}'\right|^2+j^2\left|\hat{P}\right|^2\right)-\tfrac{i}{j}R\left(aU^\prime+\zeta^\prime\right)\hat{P}^{\prime}\hat{P}^\dagger\right]\:\mathrm{d}z\geq  0 \quad \forall~\hat{P}\in\mathcal{P}.
\end{equation}

The computation of $R_o$ and $R_\varphi$ using QUINOPT then proceeds analogously to the computations described in \cref{app: background method numerical scheme derivation}. The constraints are enforced for a finite set of wavenumbers, and $\hat{P}$ and $\hat{T}$ are expanded in Legendre bases. Convergence of the polynomial degrees is checked similarly as in \cref{tbl:bounds convergence}.

\subsection{Constraining the background field so that Busse's criterion is satisfied}
\label{app: restricted background fields optimal upper bounds}

To implement the computations described in \cref{sec: restricted background profiles}, we extend the formulation~\cref{eq: UB opt cons 2} to include the additional constraints~\cref{eq: new constrained problem}. Like the first two constraints \cref{eq: k=0 constraint,eq: k>0 constraints}, the additional constraints can be enforced separately for each wavenumber. The first constraint in~\cref{eq: UB opt cons 2} becomes
\begin{multline}
\label{eq: Ro as constraint}
(a-1)\int_{-\frac{1}{2}}^{\frac{1}{2}}\left[\beta\left(\tfrac{1}{k^{2}}\left|\hat{P}''\right|^2+2\left|\hat{P}'\right|^2+k^2\left|\hat{P}\right|^{2}\right)+\tfrac{1}{k^2}\left|\hat{T}^{\prime}\right|^2+\left|\hat{T}\right|^{2}\right]\mathrm{d}z \\
	-\tfrac{i}{k}\Rey \int_{-1 / 2}^{1 / 2} \left(aU^\prime+\zeta^\prime\right) \hat{P} \hat{T}^\dagger\,\mathrm{d}z\geq 0\quad\forall\hat{P}\in\mathcal{P}_o,\:\hat{T}\in\mathcal{T}_o,
\end{multline}
where $\hat{P}(z)$ and $\hat{T}(z)$ are Fourier coefficients, and $\mathcal P_o$ and $\mathcal T_o$ contain odd functions satisfying the boundary conditions, as in \cref{app: minimizing restricted functionals}. The second constraint in~\cref{eq: UB opt cons 2} becomes
\begin{multline}
(a-1)\int_{-\frac{1}{2}}^{\frac{1}{2}}\left[
\left(\tfrac{1}{j^2}\left|\hat{P}''_e\right|^2+2\left|\hat{P}'_e\right|^2+j^2\left|\hat{P}_e\right|^2\right)+(1-\beta)\left(\tfrac{1}{j^2}\left|\hat{P}''_o\right|^2+2\left|\hat{P}'_o\right|^2+j^2\left|\hat{P}_o\right|^2\right)
\right]\mathrm{d}z \\
\label{eq: Rphi as constraint}
-\tfrac{i}{j}\Rey\int_{-\frac{1}{2}}^{\frac{1}{2}}\left(aU^\prime+\zeta^\prime\right)\left(\hat{P}_o^\prime\hat{P}_e^\dagger-\hat{P}_e^\prime\hat{P}_o^\dagger\right)\mathrm{d}z\geq 0\quad\forall\hat{P}\in\mathcal{P}.
\end{multline}
where subscripts denote even and odd parts of $\hat{P}$. In the case of upper bounds, the resulting optimization problem is
\begin{equation}
\overline{\left\langle|\nabla\vu|^2\right\rangle}\leq\min_{\substack{a,\,\zeta,\beta}}B
\quad\text{s.t.}
\begin{array}[t]{l} \cref{eq: k=0 constraint}, \\
\cref{eq: k>0 constraints}~\forall k\in K, \\
\cref{eq: Ro as constraint}~\forall k\in K, \\
\cref{eq: Rphi as constraint}~\forall j\in J,
\end{array}
\label{eq: UB chi as constraint}
\end{equation}
where $\beta\in(0,1)$, and $J$ is the set of streamwise wavenumbers for which we enforce the last constraint. We implemented~\cref{eq: UB chi as constraint} in QUINOPT to produce the bounds plotted as hollow symbols in \cref{fig: couette}. Minimizing over $a$ and $\beta$ simultaneously is a non-convex problem because they multiply each other in constraints \cref{eq: Ro as constraint,eq: Rphi as constraint}; nonetheless, the global minimum of~\cref{eq: UB chi as constraint} can be found. The optimization problem that QUINOPT formulates can be solved using the branch-and-bound algorithm implemented in YALMIP's bmibnb solver, using Mosek to solve a sequence of SDPs.

\section{Proof of Busse's theorem}
\label{app: proof of busse theorem}

This appendix gives an expository proof of the theorem from \citet{busse_property_1972} that is stated above in~\cref{eq: busse criterion}, which concerns the energy stability problem for planar shear flow models with the symmetry \eqref{eq: sym}. For the minimization problem giving the critical $\Rey$ of energy stability, \cref{app: sub functional derivation} derives a lower bound in terms of three subsidiary minimizations. \Cref{app: euler lagrange} then shows that the subsidiary minima are unchanged if taken over certain \mbox{2.5-D} or \mbox{2-D} velocity fields.

\subsection{Lower bound by three subsidiary minimizations}
\label{app: sub functional derivation}

We first decompose the poloidal and toroidal potentials into parts that are even and odd in the wall-normal coordinate $z$, denoted $\varphi=\varphi_e+\varphi_o$ and $\psi=\psi_e+\psi_o$, where
\begin{equation}
	\varphi_e(x,y,z)=\frac{1}{2}\left[\varphi(x,y,z)+\varphi(x,y,-z)\right],\quad
	\varphi_o(x,y,z)=\frac{1}{2}\left[\varphi(x,y,z)-\varphi(x,y,-z)\right]
\end{equation}
and likewise for $\psi$. The critical $\Rey$ of energy stability is the minimum of the $\funcr$ functional over mean-free velocity fields, where expression~\cref{eq: R pol tor} gives $\funcr$ in terms of $\varphi$ and $\psi$. We let $N$ and $D$ denote the numerator and denominator of $\funcr$. After the even--odd decompositions,
\begin{equation}
\label{s func tor pol odd even}
\funcr[\vw]=\frac{N}{ D}=\frac{\left\langle|\vzhat\times\nabla\lap\varphi_e|^2\right\rangle+\left\langle|\vzhat\times\nabla\lap\varphi_o|^2\right\rangle+\left\langle|\ccurl(\psi_o\vzhat)|^2\right\rangle+\left\langle|\ccurl(\psi_e\vzhat)|^2\right\rangle}
{\left|\left\langle h\left(-\lap_2\varphi_o\right)\left(\partial_y\psi_o+\partial_{xz}\varphi_e\right)\right\rangle+\left\langle h\left(-\lap_2\varphi_e\right)\left(\partial_y\psi_e+\partial_{xz}\varphi_o\right)\right\rangle\right|},
\end{equation}
where $h(z)$ is even for models with the symmetry~\cref{eq: sym}. (In the energy stability problem, $h=U'$ by definition.) Noting that $\left\langle|\vzhat\times\nabla\lap\varphi|^2\right\rangle=\left\langle|\partial_y\lap\varphi|^2\right\rangle+\left\langle|\partial_x\lap\varphi|^2\right\rangle$, we expand the numerator of $\funcr$ as
\begin{equation}
	\label{numerator labeling}
	\begin{split}
		N=&\overbrace{\langle|\partial_y\lap\varphi_e|^2\rangle+\langle|\ccurl\psi_e\vzhat|^2\rangle}^{N_e}+\overbrace{\beta \langle|\partial_y\lap\varphi_o|^2\rangle+\langle|\ccurl\psi_o\vzhat|^2\rangle}^{N_o} \\  &+\underbrace{\langle|\partial_x\lap\varphi_e|^2\rangle+\left(1-\beta\right) \langle|\partial_x\lap\varphi_o|^2\rangle+\left(1-\beta\right) \langle|\partial_y\lap\varphi_o|^2\rangle}_{N_\varphi}+\beta \langle|\partial_x\lap\varphi_o|^2\rangle,
	\end{split}
\end{equation}
where a parameter $\beta\in(0,1)$ has been introduced to split some terms, and $N_e,N_o$ and $N_\varphi$ are introduced to group certain terms. The last term in~\cref{numerator labeling} is non-negative, so $N\ge N_e+N_o+N_\varphi$. Using the triangle inequality to separate terms in the denominator of $\funcr$ gives the upper bound
\begin{equation}
	\label{triag inequality for lower bound}
	D\leq\underbrace{|\langle h\lap_2\varphi_e\partial_y\psi_e\rangle|}_{D_e}+\underbrace{|\langle h\lap_2\varphi_o\partial_y\psi_o\rangle|}_{D_o}+\underbrace{|\langle h\lap_2\varphi_e\partial_{xz}\varphi_o\rangle+\langle h\lap_2\varphi_o\partial_{xz}\varphi_e\rangle|}_{D_\varphi}.
\end{equation}
The groupings of terms are such that $N_e,D_e$ are functionals of even potentials, $N_o,D_o$ are functionals of odd potentials, and $N_\varphi,D_\varphi$ are functionals of the poloidal part. The lower bound on $N$ and upper bound on $D$ together give
\begin{equation}
	\label{string of lower bounds}
	\funcr[\vw]\geq\frac{N_e+N_o+N_\varphi}{D_e+ D_o+D_\varphi}.
\end{equation}
For any positive values of the terms in the numerator and denominator,
\begin{equation}
	\label{numbers lemma}
	\frac{N_e+N_o+N_\varphi}{D_e+ D_o+D_\varphi}\geq\min\left\{\frac{N_e}{D_e},\frac{N_o}{D_o},\frac{N_\varphi}{D_\varphi}\right\}.
\end{equation}
Therefore, the minimum of $\funcr$ can be bounded below by
\begin{align}
	\min_{\vw\in\dot{\mathcal{H}}_{3D}}\funcr[\vw]
	& \geq \min_{\vw\in\dot{\mathcal{H}}_{3D}} \min\left\{\frac{N_e}{D_e},\frac{N_o}{D_o},\frac{N_\varphi}{D_\varphi}\right\} \\
	&=\min\left\{\min_{\vw\in\dot{\mathcal{H}}_{3D}} \frac{N_e}{D_e},\ \min_{\vw\in\dot{\mathcal{H}}_{3D}} \frac{N_o}{D_o},\ \min_{\vw\in\dot{\mathcal{H}}_{3D}} \frac{N_\varphi}{D_\varphi}\right\} \\
	&=\min\left\{\min_{\substack{\vw\in\dot{\mathcal{H}}_{3D}\\\varphi,\psi\text{ even}}} \frac{N_e}{D_e},\
	\min_{\substack{\vw\in\dot{\mathcal{H}}_{3D}\\\varphi,\psi\text{ odd}}} \frac{N_o}{D_o},\
	\min_{\substack{\vw\in\dot{\mathcal{H}}_{3D}\\\psi=0}} \frac{N_\varphi}{D_\varphi}\right\},
	\label{eq: 3 mins 3D}
\end{align}
where the last equality follows because each ratio $N_e/D_e,\:N_o/D_o,\: N_\varphi/D_\varphi$ depends only on certain terms in the even--odd decompositions of $\varphi$ and $\psi$.

The utility of the lower bound~\cref{eq: 3 mins 3D} is that each right-hand minimization can be shown to admit \mbox{2.5-D} or \mbox{2-D} minimizers using arguments that are not directly applicable to the left-hand minimization. We give these dimension reduction arguments below in \cref{app: euler lagrange}. The first right-hand minimum in~\cref{eq: 3 mins 3D} is the same over \mbox{2.5-D} fields, so
\beq
\min_{\substack{\vw\in\dot{\mathcal{H}}_{3D}\\\varphi,\psi\text{ even}}} \frac{N_e}{D_e}
= \min_{\substack{\vw\in\dot{\mathcal{H}}_{2.5D}\\\varphi,\psi\text{ even}}} \frac{N_e}{D_e}
=\min_{\substack{\vw\in\dot{\mathcal{H}}_{2.5D}\\\varphi,\psi\text{ even}}} \funcr[\vw]
=R_e.
\label{eq: Ne/De R relation}
\eeq
The first equality in~\cref{eq: Ne/De R relation} is shown in \cref{app: euler lagrange} below, provided that flows of any spanwise period are admissible. The second equality in~\cref{eq: Ne/De R relation} follows because the $N_e/D_e$ and $\funcr$ functionals coincide on the subspace of \mbox{2.5-D} fields with even potentials, although they differ outside this subspace, and the last equality is the definition of $R_e$ in~\cref{eq: 3 mins}.

For the second right-hand minimum in~\cref{eq: 3 mins 3D}, we first make its $\beta$-dependence explicit by
\beq
\min_{\substack{\vw\in\dot{\mathcal{H}}_{3D}\\\varphi,\psi\text{ odd}}} \frac{N_o}{D_o}
= \sqrt{\beta}\min_{\substack{\vw\in\dot{\mathcal{H}}_{3D}\\\varphi,\psi\text{ odd}}} \frac{1}{\sqrt{\beta}}\frac{N_o}{D_o}
=\sqrt{\beta} \min_{\substack{\vw\in\dot{\mathcal{H}}_{3D}\\\varphi,\psi\text{ odd}}} \funcs_o[\vw],
\label{eq: No/Do R relation}
\eeq
where the $\beta$-independent definition of $\funcs_o$ is
\beq
\label{app eq: odd sub functional}
\funcs_o[\vw] = \frac{N_o}{\sqrt{\beta}D_o}\Bigg|_{\beta=1}
=\frac{\left\langle\left|\partial_{y} \lap \varphi_{o}\right|^{2}\right\rangle+\left\langle\left|\nabla \times \nabla \times (\psi_{o} \vzhat)\right|^{2}\right\rangle}
{\left|\left\langle h \lap_{2} \varphi_{o} \partial_{y} \psi_{o}\right\rangle\right|}.
\eeq
To see that the second inequality in~\cref{eq: No/Do R relation} holds, note that any value attained by $N_o/\sqrt{\beta}D_o$ with potentials $(\varphi_o,\psi_o)$ and $\beta\in(0,1]$ is also attained by $\funcs_o$ with potentials $(\sqrt{\beta}\varphi_o,\psi_o)$. Continuing from~\cref{eq: No/Do R relation}, we further find
\beq
\min_{\substack{\vw\in\dot{\mathcal{H}}_{3D}\\\varphi,\psi\text{ odd}}} \frac{N_o}{D_o}
=\sqrt{\beta} \min_{\substack{\vw\in\dot{\mathcal{H}}_{3D}\\\varphi,\psi\text{ odd}}} \funcs_o[\vw]
=\sqrt{\beta} \min_{\substack{\vw\in\dot{\mathcal{H}}_{2.5D}\\\varphi,\psi\text{ odd}}} \funcs_o[\vw]
=\sqrt{\beta} \min_{\substack{\vw\in\dot{\mathcal{H}}_{2.5D}\\\varphi,\psi\text{ odd}}} \funcr[\vw]
=\sqrt{\beta}R_o.
\label{eq: No/Do R relation 2}
\eeq
The second equality in~\cref{eq: No/Do R relation 2} is shown in \cref{app: euler lagrange}, provided that flows of any streamwise period are admissible. The third equality follows because the $N_o/D_o$ and $\funcr$ functionals coincide on the subspace of \mbox{2.5-D} fields with odd potentials, and the last equality is from the definition of $R_o$ in~\cref{eq: 3 mins}. Arguments for the minimum of $N_\varphi/D_\varphi$ in~\cref{eq: 3 mins 3D} are analogous to those for $N_o/D_o$, and they give
\begin{multline}
	\min_{\substack{\vw\in\dot{\mathcal{H}}_{3D}\\\psi=0}} \frac{N_\varphi}{D_\varphi}
	=\sqrt{1-\beta} \min_{\substack{\vw\in\dot{\mathcal{H}}_{3D}\\\psi=0}} \funcs_\varphi[\vw]
	=\sqrt{1-\beta} \min_{\substack{\vw\in\dot{\mathcal{H}}_{2D}\\\psi=0}} \funcs_\varphi[\vw] \\
	=\sqrt{1-\beta} \min_{\substack{\vw\in\dot{\mathcal{H}}_{2D}\\\psi=0}} \funcr[\vw]
	=\sqrt{1-\beta}R_\varphi,
	\label{eq: Nphi/Dphi R relation}
\end{multline}
where
\beq
\label{app eq: polo sub functional}
\funcs_\varphi[\vw]
= \frac{1}{\sqrt{1-\beta}}\frac{N_\varphi}{D_\varphi}\bigg|_{\beta=0}
=\frac{\left\langle\left|\vzhat\times\nabla\lap\varphi_o\right|^2\right\rangle+\left\langle\left|\partial_{x} \lap \varphi_{e}\right|^{2}\right\rangle}{\left|\left\langle h \lap_{2} \varphi_{e} \partial_{x z} \varphi_{o}\right\rangle+\left\langle h \lap_{2} \varphi_{o} \partial_{x z} \varphi_{e}\right\rangle\right|}.
\eeq
The first equality in~\cref{eq: Nphi/Dphi R relation} is the observation that $N_\varphi/\sqrt{1-\beta}D_\varphi$ has the same minimum for all $\beta\in[0,1)$. The second equality is shown in \cref{app: euler lagrange}. The third equality holds because $\funcs_\varphi$ and $\funcr$ coincide on the subspace of \mbox{2-D} poloidal fields, and the last is from the definition of $R_\varphi$ in~\cref{eq: 3 mins}.

All three right-hand minima in~\cref{eq: 3 mins 3D} have now been expressed as minimizations of the original $\funcr$ functional over certain \mbox{2.5-D} or \mbox{2-D} subspaces. By expressing the right-hand side of~\cref{eq: 3 mins 3D} in terms of $R_e$, $R_o$ and $R_\varphi$ according to \cref{eq: Ne/De R relation,eq: No/Do R relation 2,eq: Nphi/Dphi R relation}, we obtain the key lower bound~\cref{eq: pre main result} claimed in \cref{sec: busse result}. With that bound established, the paragraph surrounding~\cref{eq: pre main result} completes the proof of Busse's criterion~\cref{eq: busse criterion}.

It remains only to justify the steps in~\cref{eq: Ne/De R relation,eq: No/Do R relation 2} restricting to \mbox{2.5-D} fields, and in~\cref{eq: Nphi/Dphi R relation} restricting to \mbox{2-D} fields. The next subsection gives these arguments in terms of the $\beta$-independent functionals $\funcs_o$ and $\funcs_\varphi$, but we observe that the same arguments hold for the $\beta$-dependent functionals $N_o/D_o$ and $N_\varphi/D_\varphi$. This observation is needed in \cref{sec: restricted background profiles} to justify that the constraints in~\cref{eq: new constrained problem} are enforced only for \mbox{2.5-D} and \mbox{2-D} velocity fields, respectively.

\subsection{Symmetries of optimizers to the subsidiary problems}
\label{app: euler lagrange}

It remains to justify the steps in~\cref{eq: Ne/De R relation,eq: No/Do R relation 2,eq: Nphi/Dphi R relation} where minimizations over \mbox{3-D} fields are restricted to \mbox{2.5-D} or \mbox{2-D} fields without changing the minima. In~\cref{eq: Ne/De R relation}, it is the first equality that must be justified. For the left-hand minimization in~\cref{eq: Ne/De R relation}, which is over \mbox{3-D} fields with $\varphi$ and $\psi$ even in $z$, the Euler--Lagrange equations of $N_e/D_e$ are
\begin{equation}
\label{EL for s functional even vector form}
\nabla^4P+\tfrac{1}{2} \widetilde{R} h\lap_2T=0, \quad 
\lap\lap_2T+\tfrac{1}{2}\widetilde{R} h\lap_2P=0,
\end{equation}
where $P=-\partial_y^2\varphi_e$ and $T=\partial_y\psi_e$, and the minimum eigenvalue $\widetilde{R}$ gives the minimum of the \mbox{3-D} variational problem. Fourier transforming in the periodic $x$ and $y$ directions gives
\begin{equation}
\label{EL for s functional even fourier}
\Big(\tfrac{\rm d^2}{{\rm d}z^2}-|\mathbf{k}|^2\Big)^2\hat{P}+\tfrac{1}{2}\widetilde{R} h|\mathbf{k}|^2\hat{T}=0, \quad
|\mathbf{k}|^2\Big(\tfrac{\rm d^2}{{\rm d}z^2}-|\mathbf{k}|^2\Big)\hat{T}+\tfrac{1}{2}\widetilde{R} h|\mathbf{k}|^2\hat{P}=0
\end{equation}
for each wavevector $\mathbf{k}=(j,k)$. The minimum eigenvalue of~\cref{EL for s functional even fourier}, minimized over admissible $\mathbf{k}$, gives the minimum of the \mbox{3-D} variational problem in~\cref{eq: Ne/De R relation}. The key observation is that~\cref{EL for s functional even fourier} has the same eigenvalues for all $\mathbf{k}$ with the same magnitude. In particular, if $\widetilde{R}$ is an eigenvalue of~\cref{EL for s functional even fourier} for some wavevector $(j,k)$, it is also an eigenvalue for the wavevector $(0,\sqrt{j^2+k^2})$. The latter corresponds to a \mbox{2.5-D} eigenfunction, so the left-hand minimum in~\cref{eq: Ne/De R relation} is attained by \mbox{2.5-D} fields, meaning that the first equality in~\cref{eq: Ne/De R relation} is justified. Note that similar reasoning does not imply the existence of \mbox{2-D} eigenfunctions with wavevectors $(\sqrt{j^2+k^2},0)$; these would correspond to zero velocity fields since $P$ and $T$ are defined as $y$-derivatives of $\varphi$ and $\psi$. Note also that this reasoning requires the spanwise wavenumber $\sqrt{j^2+k^2}$ to be admissible, so the conclusion does not necessarily apply if one fixes a spanwise period of the domain.

In~\cref{eq: No/Do R relation 2}, it is the second equality that must be justified. However, the Euler--Lagrange equations of the second minimization in~\cref{eq: Ne/De R relation} are the same as~\cref{EL for s functional even vector form}. Thus, the argument following~\cref{EL for s functional even vector form} applies identically, and the second equality in~\cref{eq: No/Do R relation 2} is justified.

In~\cref{eq: Nphi/Dphi R relation}, it is the second equality that must be justified. For the second minimization in~\cref{eq: Nphi/Dphi R relation}, which is over \mbox{3-D} fields with $\psi=0$, the Euler--Lagrange equations of $\funcs_\varphi$ are
\begin{equation}
\label{EL for s functional varphi}
\begin{split}
\nabla^4\partial_xP_e-\tfrac{1}{2}\widetilde{R}\big[\partial_z\big(h\partial_xP_o\big)+h\partial_{zx}P_o\big]&=0, \\  
\nabla^4P_o-\tfrac{1}{2}\widetilde{R}\left[\partial_z\big(h\lap_2P_e\big)+h\partial_z\lap_2P_e\right]&=0,
\end{split}
\end{equation}
where $P_e=\partial_x\varphi_e$ and $P_o=\lap_2\varphi_o$. Fourier transforming in the periodic directions gives
\begin{equation}
\label{EL for s functional varphi fourier}
\begin{split}
\Big(\tfrac{\rm d^2}{{\rm d}z^2}-|\mathbf{k}|^2\Big)^2\hat{P}_e-\tfrac{1}{2}\widetilde{R}\left(h'\hat{P}_o+2h\hat{P}_o'\right)&=0, \\  
\left(\tfrac{\rm d^2}{{\rm d}z^2}-|\mathbf{k}|^2\right)^2\hat{P}_o-\tfrac{1}{2}|\mathbf k|^2\widetilde{R}\left(h'\hat{P}_e+2h\hat{P}_e'\right)&=0,
\end{split}
\end{equation}
where primes denote $\tfrac{\rm d}{{\rm d}z}$. The minimum eigenvalue $\widetilde{R}$ among admissible wavevectors $(j,k)$ is equal to the minimum of $\funcs_\varphi$ in the \mbox{3-D} variational problem. As with the eigenproblem~\cref{EL for s functional even fourier} above, the problem~\cref{EL for s functional varphi fourier} has the same eigenvalues for all $\mathbf{k}$ with the same magnitude. However, whereas~\cref{EL for s functional even fourier} does not admit spanwise-invariant eigenfunctions because $P$ and $T$ are $y$-derivatives of $\varphi$ and $\psi$, here,~\cref{EL for s functional varphi fourier} does not admit streamwise-invariant eigenfunctions because $P_e$ and $P_o$ are $x$-derivatives of $\varphi_e$ and $\varphi_o$. Thus, we conclude that if $\widetilde{R}$ is an eigenvalue of~\cref{EL for s functional varphi fourier} for some wavevector $(j,k)$, it is also an eigenvalue for the wavevector $(\sqrt{j^2+k^2},0)$. The latter wavevector corresponds to an eigenfunction that is spanwise-invariant and purely poloidal, and therefore is \mbox{2-D}, which justifies the second equality in~\cref{eq: Nphi/Dphi R relation}. This reasoning requires the streamwise wavenumber $\sqrt{j^2+k^2}$ to be admissible, so the conclusion does not necessarily apply if one fixes a streamwise period of the domain. The proof of Busse's theorem is now complete.

\bibliographystyle{jfm}
\bibliography{references}

\end{document}